\documentclass[10pt,journal,compsoc]{IEEEtran}

\ifCLASSOPTIONcompsoc
  \usepackage[nocompress]{cite}
\else
\fi

\ifCLASSINFOpdf
\else
\fi

\usepackage{graphicx}
\usepackage{graphics}
\usepackage{longtable}
\usepackage{bm}
\usepackage{algorithm}
\usepackage{algorithmic}
\usepackage{color}
\usepackage{cite}
\usepackage{amsmath}

\begin{document}

\title{Adversarial Deep Learning for Over-the-Air Spectrum Poisoning Attacks}

\author{Yalin~E.~Sagduyu,%~\IEEEmembership{Senior Member,~IEEE,}
        Yi~Shi, %~\IEEEmembership{Senior Member,~IEEE,}
        and~Tugba~Erpek%~\IEEEmembership{Student Member,~IEEE}% <-this % stops a space
\IEEEcompsocitemizethanks{\IEEEcompsocthanksitem Y. E. Sagduyu is with Intelligent Automation, Inc., Rockville, MD, USA.\protect\\
E-mail: ysagduyu@i-a-i.com
\IEEEcompsocthanksitem Y. Shi and T. Erpek are with Virginia Tech., Electrical and Computer Engineering, Blacksburg/Arlington, VA, USA. Email: \{yshi,terpek\}@vt.edu
}% <-this % stops an unwanted space
\thanks{\textsuperscript{\textcopyright} 2019 IEEE. Personal use of this material is permitted. Permission from IEEE must be obtained for all other uses, in any current or future media, including reprinting/republishing this material for advertising or promotional purposes, creating new collective works, for resale or redistribution to servers or lists, or reuse of any copyrighted component of this work in other works.}
\vspace{0.5cm}}

% The paper headers
%\markboth{Transactions on Mobile Computing,~Vol.~??, No.~?, Month~Year}%
%{Sagduyu \MakeLowercase{\textit{et al.}}: Adversarial Deep Learning for Over-the-Air Spectrum Poisoning Attacks}

\IEEEtitleabstractindextext{%
\begin{abstract}
An adversarial deep learning approach is presented to launch over-the-air spectrum poisoning attacks. A transmitter applies deep learning on its spectrum sensing results to predict idle time slots for data transmission. In the meantime, an adversary learns the transmitter's behavior (exploratory attack) by building another deep neural network to predict when transmissions will succeed. The adversary falsifies (poisons) the transmitter's spectrum sensing data over the air by transmitting during the short spectrum sensing period of the transmitter. Depending on whether the transmitter uses the sensing results as test data to make transmit decisions or as training data to retrain its deep neural network, either it is fooled into making incorrect decisions (evasion attack), or the transmitter's algorithm is retrained incorrectly for future decisions (causative attack). Both attacks are energy efficient and hard to detect (stealth) compared to jamming the long data transmission period, and substantially reduce the throughput. A dynamic defense is designed for the transmitter that deliberately makes a small number of incorrect transmissions (selected by the confidence score on channel classification) to manipulate the adversary's training data. This defense effectively fools the adversary (if any) and helps the transmitter sustain its throughput with or without an adversary present.
\end{abstract}

}

% make the title area
\maketitle

\IEEEpeerreviewmaketitle

\IEEEraisesectionheading{\section{Introduction}\label{sec:introduction}}

\IEEEPARstart{M}{achine} learning
provides wireless communications 
with automated means to learn from and adapt to dynamic spectrum environment that includes a variety of topology, channel, traffic, and interference effects \cite{Clancy2007,Chen2017,Simeone}. Examples of machine learning applications in wireless communications 
include spectrum sensing \cite{Lee2017}, channel estimation \cite{DeepOFDM}, spectrum access \cite{Yi2018}, power control \cite{Tugba2018}, signal classification \cite{OShea2016}, and augmentation \cite{Kemal2018}.

Wireless communication is vulnerable to different types of \emph{attacks} such as jamming \cite{Trappe} and eavesdropping \cite{Zou15:eavesdropping} due to its open broadcast nature. 
Dynamic spectrum access (DSA) is especially sensitive to attacks as it involves various tunable parameters that can be manipulated by adversaries
\cite{Clancy08:CogSec}. One example is the primary user emulation (PUE) attack, where an adversary pretends to be a primary user and aims to decrease the spectrum access opportunities of cognitive radios \cite{PUE}. In a collaborative sensing environment, another example is the \emph{spectrum sensing data falsification} (SSDF) attack that targets the spectrum sensing operation by falsifying spectrum sensing reports \cite{Bian2008}.

As machine learning starts finding more applications in wireless communications, the safe use of machine learning algorithms is emerging as a major security concern. In particular, machine learning itself may become the target of the adversary. Such security issues have been studied in other data domains (e.g., computer vision) in the emerging field of \emph{adversarial machine learning}. Examples include \emph{exploratory (inference) attacks} to infer how a machine learning algorithm operates (e.g., learn a classifier's decision boundaries) \cite{Fredrikson}, 
\emph{evasion attacks} to fool a machine learning algorithm into making wrong decisions (e.g., fool a trained filter into misclassifying spam emails) \cite{Kurakin,Biggio}, and \emph{causative} attacks to provide incorrect information (e.g., training data in supervised learning) to a machine learning algorithm \cite{Biggio2}.

When adversarial machine learning is applied to wireless communications, the objective is not anymore to directly attack wireless communications but to manipulate the underlying cognitive engine based on machine learning algorithms. One important difference from other data domains (e.g., computer vision classifier APIs) is that the adversary and the target in wireless communications observe different features (due to different channel and interference effects) and use different classification labels (as they perceive different events). In \cite{Yi2018,Tugba2018},  an adversary was designed to build a deep neural network to mimic how the DSA algorithm works and then jams the data transmissions by first running its own surrogate classifier to determine successful transmission opportunities.
While this attack decreases the throughput, it incurs major energy consumption and leaves a large footprint for easy detection.

In this paper, we develop a new type of wireless attack based on adversarial machine learning, namely the \emph{over-the-air spectrum poisoning attack} that targets the sensing period of a transmitter under attack.
This is a stealth attack that is energy-efficient and does not leave a large footprint compared with previous attacks.
Unlike traditional denial of service attacks \cite{DOS}, where the adversary transmits to jam data transmissions,
the adversary aims to manipulate the spectrum sensing data by jamming the spectrum sensing period so that the target transmitter makes wrong decisions by using the unreliable spectrum sensing results.
This attack also differs from the SSDF attack, since the adversary does not participate in cooperative spectrum sensing and does not try to change the estimated channel labels directly as in the SSDF attack.
Instead, the adversary injects adversarial perturbations over the air (in terms of jamming the spectrum sensing period) to the channel in order to fool the transmitter into making wrong transmit decisions or make the transmitter's re-training process fail.
To counteract such attacks,  we develop a \emph{defense mechanism} that uses the classification outputs of the transmitter's deep neural network to add controlled errors into channel access decisions of the transmitter and consequently mislead the adversary.

We consider a canonical wireless communication scenario with a transmitter, its corresponding receiver, an adversary, and some other background traffic. We apply different channel models including Gaussian, Rayleigh, Rician, and log-normal channels.
Note that the proposed attacks are independent of the network topology and can be directly applied to other network topologies.
We also show results for multiple background traffic sources.
For this case, aggregate traffic is observed through spectrum sensing that inputs the aggregated signal to the transmitter and jammer algorithms.
If there are multiple adversaries, each of them and the jammer can individually apply their proposed algorithms (while treating interference the same way as background traffic).
The transmitter builds a machine learning model (based on a deep neural network) to predict the busy and idle states of the channel.
The adversary applies \emph{adversarial deep learning} to launch various attacks, including \emph{exploratory attack, evasion attack in test phase, causative attack in training phase}, and their combinations. As a defense strategy, the transmitter launches an attack back on the cognitive engine of the adversary and aims to degrade the inference stage of the adversary.

The main contributions of this paper are on stealth and energy-efficient attacks on wireless communications built upon adversarial machine learning and a corresponding defense scheme. We present novel techniques for
\begin{itemize}
    \item exploratory (inference) attacks by observing the spectrum and feedback on transmission outcomes in communications (see Section~\ref{subsec:exploratory}),
    \item evasion attacks on spectrum sensing of wireless communications in test phase (see Section~\ref{subsec:evasion}),
       \item causative attacks on spectrum sensing of wireless communication in training phase (see Section~\ref{subsec:causative}), and
       \item defense scheme against all these attacks (see Section~\ref{subsec:defense}).
\end{itemize} 

\subsection{Exploratory (Inference) Attack}
\label{subsec:exploratory}

The adversary applies \emph{adversarial deep learning} to launch an \emph{exploratory attack}.
For that purpose, the adversary trains a deep neural network.
Subsequently, it intentionally changes the transmitter's sensing results by transmitting when it predicts that there will be a successful transmission if there were no attacks.

The training data of the transmitter consists of  time-series of spectrum sensing results as features and channel idle/busy status based on the ground truth (the background transmitter's on/off state) as labels. Using this training data, the transmitter builds a deep neural network to make transmit decisions.
If a transmission is successful (i.e., the signal-to-interference-plus-noise ratio (SINR) exceeds a threshold), the receiver sends an acknowledgement (ACK) to the transmitter and the adversary can detect the presence of this ACK (without decoding its content).

The adversary first determines the time slot structure used by the transmitter and then performs an \emph{exploratory attack} to build a classifier that can predict the outcome of transmissions, i.e., whether there will be an ACK if there is no attack.
The adversary uses time-series of its own spectrum sensing results as \emph{features} and presence/absence of ACKs as \emph{labels} in its training and test data.
Note that this is not a standard exploratory attack and the classifier built by the adversary will not be the same as (or similar to) the classifier used by the transmitter, due to the following two differences.
\begin{itemize}
	\item The transmitter and the adversary are in different locations and thus their sensing results will vary based on the channel environment and differ from each other. As a result, the input data (features) to their classifiers will differ.
	\item The adversary predicts the outcome of the transmissions (`ACK' or `no ACK') while the transmitter predicts channel status (`idle' or `busy'). As a result, the output data of their classifiers will differ.
\end{itemize}

Once the adversary develops its deep neural network model as part of an exploratory attack, it uses this classifier to perform either an evasion attack in the test phase or causative attack in the training phase.

\subsection{Evasion Attack in Test Phase}
\label{subsec:evasion}

After building its classifier, the adversary predicts when the transmitter will have a successful transmission (if there was no attack) and performs the \emph{evasion attack} in test phase, i.e., the adversary transmits to change the channel status in order to poison (i.e., falsify) the transmitter's input (spectrum sensing data) to the machine learning algorithm.
The attack considered in this paper is similar to that in \cite{Yi2018,Tugba2018},
where the adversary also first learns the transmitter's behavior (ACK or not) by an exploratory attack and then performs subsequent attacks.
The difference is that in \cite{Yi2018,Tugba2018},
the adversary performs a standard jamming attack during the data transmission period to make a transmission fail while in this paper the adversary performs an evasion attack in the sensing period such that the transmitter is provided with incorrect input data (manipulated over the air) to its classifier and makes the wrong decision of not transmitting.
This attack is harder to detect since it does not directly jam the transmitter's signal but it changes the input data to the decision mechanism.
Moreover, this attack is more energy efficient since the adversary makes a very short transmission in the sensing period.

We show that this adversarial deep learning approach significantly reduces the transmitter's performance.
In particular, for the scenario studied in numerical results, only few transmission attempts are made when the evasion attack is launched and the achieved throughput (normalized by the best throughput by an ideal algorithm to detect every idle channel) drops from $98.96\%$ to $3.13\%$.
For comparison purposes, we consider the same energy budget (namely, the energy consumption of spectrum poisoning attack) and study an attack that jams data transmission period (much longer than spectrum sensing period).
Due to this small energy budget, an adversary cannot jam the data transmission period of all time slots that will have a successful transmission.
Thus, the normalized throughput can only be reduced to $41.67\%$, which is much higher than $3.13\%$ under the evasion attack on the data sensing period.

\subsection{Causative Attack in Training Phase}
\label{subsec:causative}

For the case that the transmitter collects additional training data and retrains its classifier, the adversary can also apply a \emph{causative attack} on the training data after determining the start and end of training phase.
The classifier of the transmitter is then updated by using some incorrect data and thus becomes worse than before retraining.
As a result, even if there is no further attack, the transmitter will make incorrect decisions in the future and its performance will drop.
This attack is even harder to detect and more energy efficient than evasion attack since the adversary's transmissions are limited to sensing periods of only the retraining phase.
For the scenario studied in numerical results, the normalized throughput drops from $98.96\%$ to $87.27\%$ while a defender that monitors data transmissions cannot find a jamming signal.
A causative attack can be followed by other attacks after the transmitter's classifier is updated, such as evasion attack on the data sensing period and jamming attack on the transmission period.
Both attacks can further reduce the transmitter's throughput.
When combined with causative attack, evasion attack reduces throughput to a smaller value ($2.72\%$) than jamming ($37.27\%$).

\begin{figure}
	\centering
	\includegraphics[width=0.7\columnwidth]{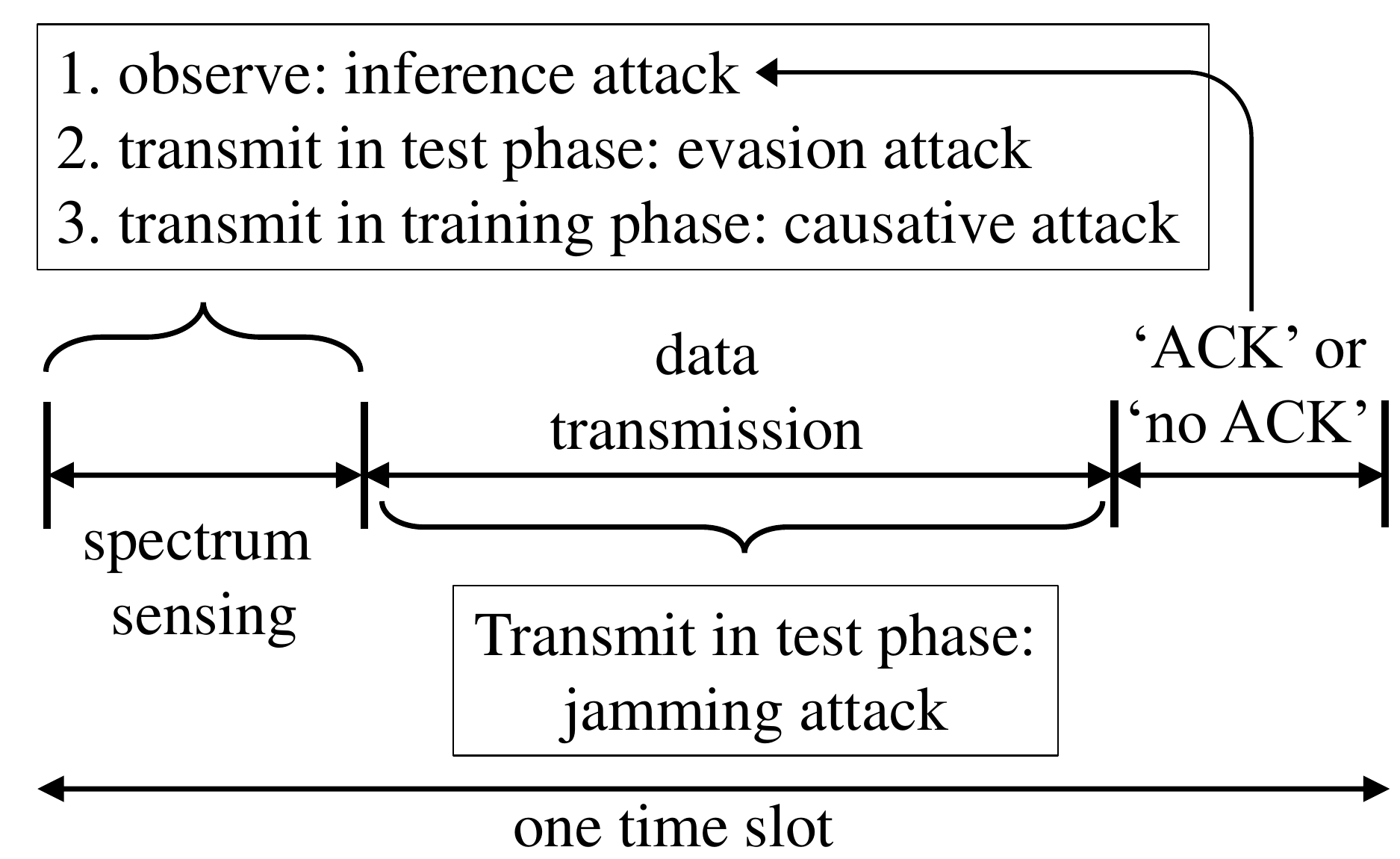} 
	\caption{Overview of  adversarial machine learning proposed in this paper.}
	\label{fig:overview}
\end{figure}

The overview of these attacks is shown in Figure~\ref{fig:overview}.

\subsection{Defense Scheme}
\label{subsec:defense}

Since the adversary can significantly reduce the transmitter's throughput, it is necessary to develop a \emph{defense} scheme.
One approach to protect a neural network against adversarial machine learning is adding randomness to the neural network structure (namely, weights and biases) \cite{Kurakin18:defense}.
This defense is effective if an adversary can access the output layer of a neural network (labels and scores). However, it cannot be applied in wireless communications, since the adversary collects its training data indirectly by observing the outcome of transmissions (without obtaining a score).
Also, small randomness may not change the outcome of transmissions while large randomness will impact $T$'s performance.
Instead, we design a defense scheme where the transmitter intentionally makes some incorrect transmit decisions to manipulate the training data of the adversary so that the adversary cannot build a reliable deep learning model. This corresponds to a causative attack by the transmitter on the adversary's inference attack stage.
These incorrect transmit decisions should be made on a carefully selected set of time slots to balance the trade-off between the large impact on the adversary's classifier and small loss in transmitter's performance due to incorrect transmit decisions. We select these time slots from those with the classification score (provided by the deep learning classifier of the transmitter) that is far away from the decision boundary.
We show that this defense mechanism can increase the normalized throughput from $3.13\%$ to $76.04\%$ against evasion attacks, and can be effectively applied against other attacks, as well, by adapting the level of defense without knowing whether an adversary is present, or not.

\subsection{Paper Organization}

The rest of the paper is organized as follows.
Section~\ref{sec:related} reviews related work on wireless attacks and adversarial machine learning.
Section~\ref{sec:scenario} describes the system model.
Section~\ref{sec:transmitter} describes the transmitter's algorithm and shows the performance without an attack.
Section~\ref{sec:adversary} describes the adversary's algorithm and shows the performance under different attacks.
Section~\ref{sec:defense} presents a defense mechanism and shows how it improves the performance.
Section~\ref{sec:conclusion} concludes the paper.

\section{Related Work}
\label{sec:related}
There are various security concerns regarding the safe use of machine learning algorithms. For example, if the input data to a machine learning algorithm is manipulated during the training or operation (test) time, the output will be very different compared to the expected results.
These particular security threats are addressed in the emerging field of \emph{adversarial machine learning}, which studies learning in the presence of adversaries and aims to enable safe adoption of machine learning to the emerging applications.

Attacks under adversarial machine learning are divided into three broad categories, namely \emph{exploratory (or inference) attacks}, \emph{evasion attacks}, and \emph{causative attacks}.
\begin{itemize}
\item In \emph{exploratory attacks} \cite{Ateniese,Tramer,Fredrikson}, the adversary aims to understand how the underlying machine learning works for an application (e.g., inferring sensitive and/or proprietary information).

\item In \emph{evasion attacks} \cite{Biggio,Kurakin}, the adversary attempts to fool the machine learning algorithm into making a wrong decision (e.g., fooling a security algorithm into accepting an adversary as legitimate).

\item In \emph{causative attacks} \cite{Biggio2}, the adversary provides incorrect information such as training data to machine learning.
\end{itemize}
These attacks can be launched separately or combined, i.e., causative and evasion attacks can be launched by making use of the inference results from an exploratory attack \cite{HST2018}.
For wireless applications, the evasion attack was considered in \cite{LarssonAML,Headley19,Deniz19, Silvija19} by adding adversarial perturbations to fool receivers to misclassify signal types (such as modulations). Adversarial distortions were considered in \cite{Tansu2018} to support anti-jamming by deceiving the jammer’s learning algorithms in a game-theoretic framework. Built upon exploratory attacks, deep learning was studied in \cite{Yi2018,Tugba2018}
to launch jamming attacks on data transmissions. This paper focuses on attacks during spectrum sensing of wireless communications. In \cite{Shi18:poisoning}, we performed the preliminary study on exploratory and evasion attacks on data sensing for wireless communications and corresponding defense strategies.

Apart from adversarial machine learning, there are different types of attacks on the spectrum sensing decisions studied in the literature \cite{Clancy08:CogSec,Zou15:SecurePhy,Xiao18:crowdsensing}.
In a collaborative sensing environment, some users may send falsified reports to each other or to a decision center. This corresponds to a \emph{spectrum sensing data falsification} (SSDF) attack  that aims to degrade the performance of spectrum sensing \cite{Penna, Yut}.
The attacks proposed in this paper are different from SSDF attacks, since the adversary does not participate in collaborative spectrum sensing and does not falsify estimated spectrum sensing results but rather, it transmits in the spectrum sensing period to change the inputs to spectrum classifier over the air. Another type of attack, the \emph{primary user emulation} (PUE) attack, aims decrease the spectrum access opportunities of cognitive radios. A defense technique for PUE attacks using belief propagation was studied in \cite{PUE}. Cognitive radio networks are also susceptible to conventional security threats such as \emph{jamming} \cite{Sagduyu11:jamming}, \emph{eavesdropping} \cite{Zou15:eavesdropping} and \emph{noncooperation} \cite{Sagduyu09:noncoop}.
These threats on wireless communications extend from physical layer to higher layers, e.g., attacks on routing in the network layer \cite{Lu2017} and network flow inference attacks \cite{LuCliff2017}. Wireless security finds rich applications of deep learning.
Deep learning was applied to authenticate signals \cite{Saad2018}, detect and classify jammers of different types \cite{TwoDim, Wu2017, gunes}, and control communications to mitigate jamming effects \cite{Poor2018, Xiao18, Xiao18:antijam}.
Using wireless sensors, deep learning was also used to infer private information in analogy to exploratory attacks \cite{Liang18}.

In this paper, we study adversarial machine learning attacks on spectrum sensing under a small energy budget. Following an exploratory attack, we consider an evasion attack in test phase and as a causative attack in training phase. We also consider combination of these attacks along with jamming data transmissions. Moreover, we propose a defense scheme in this paper to counteract these new types of attacks.

\section{System Model}
\label{sec:scenario}

\begin{figure}
	\centering
	\includegraphics[width=0.75\columnwidth]{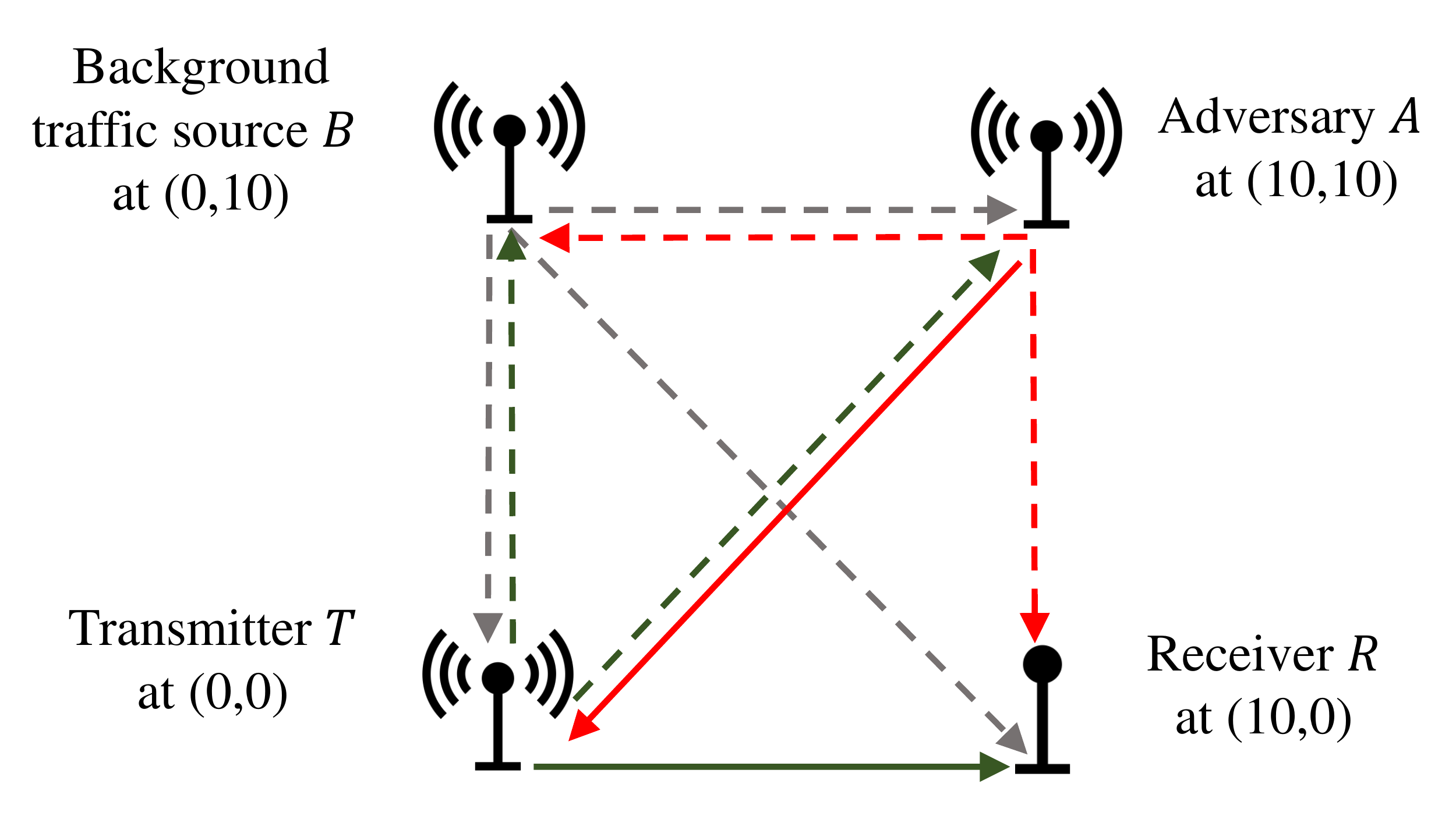} 
	\caption{The network topology.}
	\label{fig:topology}
\end{figure}

We consider a communication system that includes a transmitter $T$, a receiver $R$, an adversary $A$ and some background traffic source $B$ that may transmit its data.
These nodes operate on a single channel.
The network topology to generate numerical results is shown in Figure~\ref{fig:topology}. As noted in Section~\ref{sec:introduction}, the proposed attacks can be applied to other network topologies.
We assume that $T$ and $R$ are cognitive radios that can run algorithms developed in this paper and can perform spectrum sensing and transmit and receive data and feedback, as  specified in algorithm solution.
We mostly focus on fixed locations in this paper.
We will discuss the impact of mobile nodes in Section~\ref{sec:transmitter}.

The transmission pattern of $B$ is not known by $T$ or $A$ \emph{a priori}, and can be detected via spectrum sensing.
Packets arrive at $B$ randomly according to the Bernoulli process with rate $\lambda$ (packet/slot). If $B$ is not transmitting, it becomes active with certain probability when its queue is not empty. Once activated, it will keep transmitting until its queue becomes empty.
Thus, there may be a continuous period of busy slots and its length depends on the number of packets in $B$'s queue, which is related to the number of previous idle slots.
Therefore, channel busy/idle states are correlated over time, and both $T$ and $A$ need to observe the past channel status over a time period to predict the current channel status.

{\small
\begin{table}
	\caption{Notation.} 
	\centering
		\begin{tabular}{c|l}
			Symbol & Definition \\ \hline \hline
			$A$ & The adversary \\ \hline
			$B$ & The background traffic source \\ \hline
			$\mathcal{C}_i$ & Node $i$'s classifier \\ \hline
            $\mathcal{C}_{\tilde{T}}$ & Updated classifier for $T$ \\ \hline
			$\mathcal{C}_i(\bm{F}_t)$ & Classifier $\mathcal{C}_i$'s output on features $\bm{F}_t$ \\ \hline
			$d_{ij}$ & Distance from node $i$ to node $j$ \\ \hline
            $D_{test}$ & Test data set \\ \hline
            $\hat{D}_{test}$ & Test data set for $A$ under $T$'s defense \\
            & actions \\ \hline
            $D_{train}$ & Training data set \\ \hline
            $\hat{D}_{train}$ & Training data set for $A$ under $T$'s defense \\
            & actions \\ \hline
			$e(\mathcal{C}_i)$ & Error probability for classifier $\mathcal{C}_i$ \\ \hline
			$e_{_{FA}}(\mathcal{C}_i)$ & False alarm probability for classifier $\mathcal{C}_i$ \\ \hline
			$e_{_{MD}}(\mathcal{C}_i)$ & Misdetection probability for classifier $\mathcal{C}_i$ \\ \hline
			$\bm{F}_t$ & Features for time slot $t$ \\ \hline
			$g_{ij}(t)$ & Channel gain from node $i$ to 
$j$ at time $t$ \\ \hline
            $H$ & Set of deep learning hyperparameter \\
            & values \\ \hline
            $\mathcal{H}$ & Feasible region for deep learning \\
            & hyperparameters \\ \hline
            $l_t$ & The label (ACK or not) at time $t$ \\ \hline
            $\mathcal{L}_i(H, D_{train}, D_{test})$ &  Function for deep learning process of \\
            & node $i$ on given  $H$, $D_{train}$, and $D_{test}$ \\ \hline
			$M_{\text{Sr}}$ & Success ratio among all transmissions \\ \hline
			$M_{\textit{Th}}$ & Normalized throughput \\ \hline
			$M_{\textit{Tr}}$ & Transmission ratio \\ \hline
            $n_{busy}$ & Number of busy time slots in $D_{test}$ \\ \hline
            $n_{idle}$ & Number of idle time slots in $D_{test}$ \\ \hline
            $n_{_{FA}}(\mathcal{C}_i)$ & Number of false alarms for classifier $\mathcal{C}_i$ \\ \hline
            $n_{_{MD}}(\mathcal{C}_i)$ & Number of misdetections for classifier $\mathcal{C}_i$ j
            \\ \hline
            $n_{new}$ & Number of most recent sensing results in \\
            & each $\bm{F}_t$ \\ \hline
			$N_0$ & Noise \\ \hline
			$p_t$ & Sensed power at time $t$ \\ \hline
			$p(s)$ & Classification score for sample $s$ \\ \hline
			$P$ & Transmit power \\ \hline
            $P_d$ & Ratio of $T$'s defense actions \\ \hline
			$R$ & The receiver \\ \hline
			$s$ & A sample \\ \hline
			$S_t$ & The idle/busy status at time slot $t$ \\ \hline
            $t_i$ & An inter-arrival time between two ACKs \\ \hline
			$T$ & A transmitter \\ \hline
			$\gamma$ & SNR or SINR \\ \hline
			$\gamma_{min}$ & SNR or SINR threshold for a successful \\
& transmission \\ \hline
            $\lambda$ & Arrival rate of $B$ \\ \hline
			$\tau$ & Classification threshold
		\end{tabular}
	\label{table:notation}
\end{table}
}

Time is divided in slots. Within each slot, the initial short period of time is allocated by $T$ for spectrum sensing and the ending short period of time is allocated for feedback (i.e., ACK).
The rest of a slot is for data transmission if channel is detected as idle.
The decision of $T$ is based on a classifier $\mathcal{C}_T$ (trained by deep learning) that analyzes sensing results and then determines the time slot status such that a time slot is busy if background traffic is detected and idle otherwise.
$\mathcal{C}_T$ is independent of $A$'s actions.
Each sensing result is either
\begin{itemize}
	\item noise $N_0$ (idle time slot) or
	\item noise plus the received power from background traffic $N_0 + g_{_{BT}}(t) P$ (busy time slot), where $g_{_{BT}}(t)$ is the channel gain from $B$ to $T$ at time $t$ and $P$ is transmit power at $B$.
\end{itemize}
Data transmission is successful if the SNR $\gamma = \frac{g_{TR}(t) P_T}{N_0}$ or the SINR $\gamma = \frac{g_{TR}(t) P_T}{N_0 + g_{_{BR}}(t) P_B}$ at the receiver $R$ is not less than a threshold $\gamma_{min}$, where $g_{TR}$ is the channel gain from $T$ to $R$. 
We assume Gaussian noise at $R$ and Gaussian channel gain from $T$ to $R$.
Results for other channel models, i.e., Rayleigh channel, Rician channel, and log-normal channel, are also presented in Section~\ref{sec:transmitter}.
Channel quality changes over time.
The mean value of the channel gain is calculated based on the free-space propagation loss model.
Note that algorithms in this paper are not tied to any channel model.
$R$ sends an ACK for each successful transmission.

Before launching an attack, $A$ first determines the length of a time slot and the length of sensing, transmission, and feedback periods in a time slot based on its spectrum sensing results. 
For that purpose, $A$ senses the channel over a period of time to collect data.
Then $A$ can detect ACKs reliably because of the unique properties of ACKs. First, an ACK always follows an active data transmission period and is followed by an inactive sensing period. Second, ACK itself is a short transmission period with a reliable modulation and coding scheme, which is different than the scheme used for data transmissions.
The inter-arrival time between two ACKs is an integer times the length of a time slot, since some time slots do not have an ACK.
The problem of determining the time slot length from multiple observations of such inter-arrival times is solved by Algorithm~\ref{alg:slot}.
Once the length of a time slot is determined, $A$ can further determine the sensing and transmission periods in a time slot.
In a time slot with ACK, there is a successful transmission and thus the starting point of such a transmission (with higher sensed power than idle cases) determines the sensing period (before this point) and the data transmission period (after this point).

\begin{algorithm}[t]
    \caption{Determine the length of a time slot}
    \label{alg:slot}
    \begin{algorithmic}[1]
        \STATE $A$ observes the spectrum over a time period and identifies ACKs.
        \STATE Initialize a list $\mathcal{I}$ that includes inter-arrival times $t_1, t_2, \cdots, t_m$ of these ACKs.
        \STATE Find $t_{i^*}$ as the smallest number in $\mathcal{I}$.
        \FOR{$i=1$ to $m$, $i \neq i^*$}
            \STATE $k = \left\lfloor \frac{t_i}{t_{i^*}} \right\rfloor, \hat{t}_i = t_i - k t_{i^*}$
            \IF{$\hat{t}_i \approx 0$}
                \STATE Remove $t_i$ from $\mathcal{I}$.
            \ELSE
                \STATE Replace $t_i$ by $\hat{t}_i$.
            \ENDIF
        \ENDFOR
        \IF{$\mathcal{I}$ has only one element $t_{i^*}$}
            \STATE Return $t_{i^*}$.
        \ELSE
            \STATE Go to Step~3.
        \ENDIF
    \end{algorithmic}
\end{algorithm}

Then for each time slot, $A$ aims to predict whether there will be a successful transmission (ACK) if there is no attack.
Note that $A$ only detects the presence of the ACK message but does not need to decode it.
The prediction by $A$ is based on another classifier $\mathcal{C}_A$ that is trained by $A$ using deep learning.
If $A$ predicts that there will be a successful transmission, it performs some attack to reduce the throughput of $T$.
In this paper, we consider the attack of transmitting in the initial short sensing period to change the sensing result of $T$ for the current time slot.
Since this sensing result is an input to $\mathcal{C}_T$ on time slot status, $T$ may make a wrong decision, even if $\mathcal{C}_T$ was trained properly.

The advantage of this attack, comparing with the continuous jamming attack, is that the initial sensing period is much shorter than the data transmission period.
As a result, the power consumption of this attack is much less compared to continuous jamming.
In addition, it is harder to detect this attack compared to continuous jamming due to its small footprint.

$T$ may also apply a defense mechanism to mitigate such attacks. For that purpose, $T$ takes wrong actions in a controlled manner such that the `ACK' or `no ACK' results (namely labels for $\mathcal{C}_A$) are changed.
As a consequence, $\mathcal{C}_A$ cannot be reliably trained and the attack performance drops.
However, $T$ needs to minimize the number of these wrong actions such that the performance loss due to wrong channel access decisions remains small. In Section~\ref{sec:defense}, we will show how to carefully select a small set of time slots (depending on the classification score of $T$) and take wrong actions only in these slots to better mislead $A$.
Table~\ref{table:notation} lists the notation used in this paper.

\section{Transmitter's Algorithm}
\label{sec:transmitter}

$T$ senses the spectrum, identifies an idle time slot (when $B$ is not transmitting), and then decides whether to transmit or not.
$T$ applies a deep learning classifier $\mathcal{C}_T$ to identify idle time slots.
$\mathcal{C}_T$ is pre-trained using a number of samples, where a sample for time $t$ has the most recent $n_{new}$ sensing results $p_{t-n_{new}-1}, p_{t-n_{new}-2}, \cdots, p_t$ as features $\bm{F}_t$ and the current busy/idle status $S_t$ as the label.
$n_{new}$ is potentially a design parameter for $T$ and can be tuned by $T$ to optimize its performance. In this paper, we assume $n_{new}=10$.
Each sensing result is either a Gaussian noise with normalized power $N_0$ (idle time slot) or noise plus the transmit power from another user $N_0+g_{_{BT}}(t) P$ (busy time slot),
where noise and the channel gain are random variables with Gaussian distributions.
After observing a certain period of time, $T$ collects a number of samples to be used as training data $D_{train}$ to build a deep learning classifier $\mathcal{C}_T$.
$T$'s training algorithm is summarized
in Algorithm~\ref{alg:t-train}.

\begin{algorithm}[t]
    \caption{$T$'s training algorithm}
    \label{alg:t-train}
    \begin{algorithmic}[1]
        \STATE $T$ collects sensing data over a time period to build its training data  $D_{train}$.
        \STATE $T$ builds a training sample $\{ \bm{F}_t, S_t \}$ for each time $t \ge n_{new}$, where $\bm{F}_t = (p_{t-n_{new}-1}, p_{t-n_{new}-2}, \cdots, p_t)$, $p_t$ is the sensed power at time $t$, and $S_t$ is the busy/idle status at time $t$.
        \STATE $T$ trains a deep learning classifier $\mathcal{C}_T$ using training data $D_{train}$.
    \end{algorithmic}
\end{algorithm}

\begin{figure}[t]
	\centering
	\includegraphics[width=1\columnwidth]{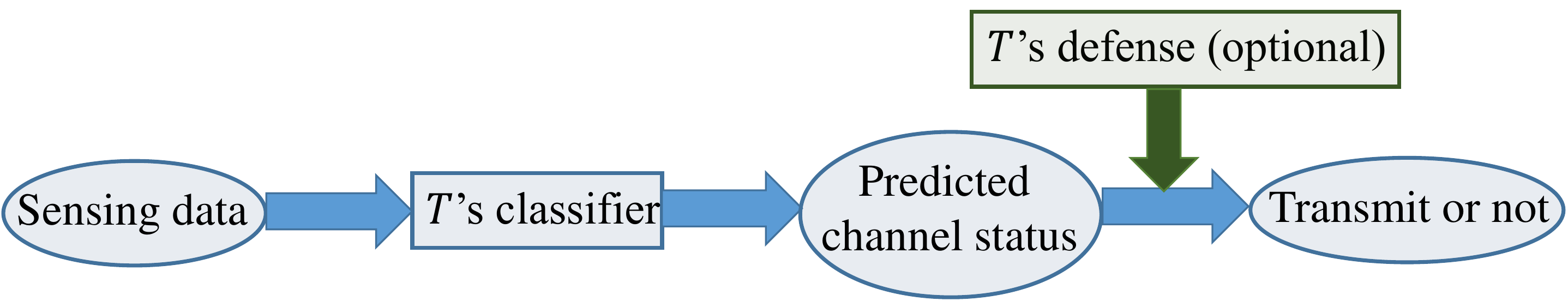} 
	\caption{$T$'s classifier in test time.}
	\label{fig:tsclassifierruntime}
\end{figure}

Once $\mathcal{C}_T$ is built, $T$ uses it to predict the channel status of each time slot and transmit if it predicts a given time slot as idle.
The block diagram in Figure~\ref{fig:tsclassifierruntime} shows $T$'s operation in test time.
Note that there is an optional block of defense,
which will be discussed later in Section~\ref{sec:defense}.
This prediction algorithm is summarized
in Algorithm~\ref{alg:t-predict}. In this algorithm, two types of errors may be incurred:
\begin{itemize}
\item \emph{Misdetection}. A busy time slot is detected as idle, i.e., $S_t =$ `busy' and $\mathcal{C}_T(\bm{F}_t)=$ `idle`.

\item \emph{False alarm}. An idle time slot it is detected as busy, i.e., $S_t =$ `idle' and $\mathcal{C}_T(\bm{F}_t)=$ `busy'.
\end{itemize}
Transmitter $T$ aims to minimize error probability $e(\mathcal{C}_T) = \max\{ e_{_{MD}}(\mathcal{C}_T), e_{_{FA}}(\mathcal{C}_T) \}$ to balance misdetections and false alarms, where $e_{_{MD}}(\mathcal{C}_T)$ is the misdetection probability and $e_{_{FA}}(\mathcal{C}_T)$ is the false alarm probability for classifier $\mathcal{C}_T$.
This objective is important, especially when data is imbalanced among labels.
These error probabilities are calculated by $e_{_{MD}}(\mathcal{C}_T) = \frac{n_{_{MD}}(\mathcal{C}_T)}{n_{busy}}$ and $e_{_{FA}}(\mathcal{C}_T) = \frac{n_{_{FA}}(\mathcal{C}_T)}{n_{idle}}$, where $n_{_{MD}}(\mathcal{C}_T)$ is the number of misdetections, $n_{busy}$ is the number of busy time slots in $D_{test}$, $n_{_{FA}}(\mathcal{C}_T)$ is the number of false alarms, and $n_{idle}$ is the number of idle time slots in $D_{test}$.
There are many hyperparameters in deep learning, e.g., the number of layers in the neural network and the number neurons per layer.
Denote $H$ as a set of hyperparameter values and $\mathcal{H}$ as the feasible region for hyperparameters.
In addition to training the deep neural network (namely, determining weights and biases), these hyperparameters should also be optimized to minimize $e(\mathcal{C}_T)$.
Hyperparameter selection leads to the following optimization problem.
\begin{eqnarray*}
\mbox{{\it OptHyper}:} \\
\mbox{minimize} && e(\mathcal{C}_T) \\
\mbox{subject to} && e(\mathcal{C}_T) \ge e_{_{MD}}(\mathcal{C}_T) , \; e(\mathcal{C}_T) \ge e_{_{FA}}(\mathcal{C}_T) , \\
&& e_{_{MD}}(\mathcal{C}_T) = \frac{n_{_{MD}}(\mathcal{C}_T)}{n_{busy}} , \; e_{_{FA}}(\mathcal{C}_T) = \frac{n_{_{FA}}(\mathcal{C}_T)}{n_{idle}} , \\
&& \mathcal{C}_T = \mathcal{L}_T(H, D_{train}) , \; H \in \mathcal{H} ,
\end{eqnarray*}
where $\mathcal{L}_T(H, D_{train})$ is $T$'s function for deep learning process on given hyperparameters $H$ and training data $D_{train}$.
Note that the closed form expression of $\mathcal{L}_T(\cdot)$ is unknown due to the complex neural network built in deep learning.
Therefore, standard optimization techniques such as convex optimization cannot be applied to solve {\it OptHyper}.
In this paper, we find local optimal solution to {\it OptHyper} by applying a greedy sequential-fixing algorithm that starts with an initial set of parameter values and optimizes one parameter value (while keeping others unchanged) in each round until all parameter values are optimized.
In addition, we solve {\it OptHyper} by Hyperband \cite{Li16}, which starts with a number of settings of parameter values and check their performance with a limited number of training epochs.
This approach can still achieve good performance (local optimal solution) with low complexity.
Based on the current performance results for each setting, some bad settings are removed.
In the next round, remaining settings will continue for more epochs and more accurate performance results will be obtained to further remove some bad settings.
After several rounds, a final solution on parameters will be obtained likely with good performance by considering many settings.
On the other hand, it has low complexity since most of settings can be removed without a complete training process.
Alternatively, a random search approach in \cite{Bergstra12:parameter} could be used for low complexity but performance is not good as only a small random portion of the large search space is covered.
{\it OptHyper} could also be solved by genetic algorithm that can find good solutions on parameters at the expense of high time complexity.\footnote{$H$ is used as chromosome and the algorithm starts with a number of initial solutions on $H$, which is the first generation. Once a termination condition (e.g., no signification improvement on $e(\mathcal{C}_T)$ over some generations) is met, the best solution in the current generation on $H$ is the final solution by the genetic algorithm.}

\begin{algorithm}[t]
    \caption{$T$'s prediction algorithm}
    \label{alg:t-predict}
    \begin{algorithmic}[1]
        \STATE At time $t$, $T$ senses channel and obtains power $p_t$.
        \STATE $T$ builds a test sample $\bm{F}_t = (p_{t-n_{new}-1}, p_{t-n_{new}-2}, \cdots, p_t)$.
        \STATE $T$ uses its classifier $\mathcal{C}_T$ to decide on a label (`busy' or `idle') for the test sample at time slot $t$, i.e., computes $\mathcal{C}_T(\bm{F}_t)$.
    \end{algorithmic}
\end{algorithm}

\begin{figure}
	\centering
	\includegraphics[width=1\columnwidth]{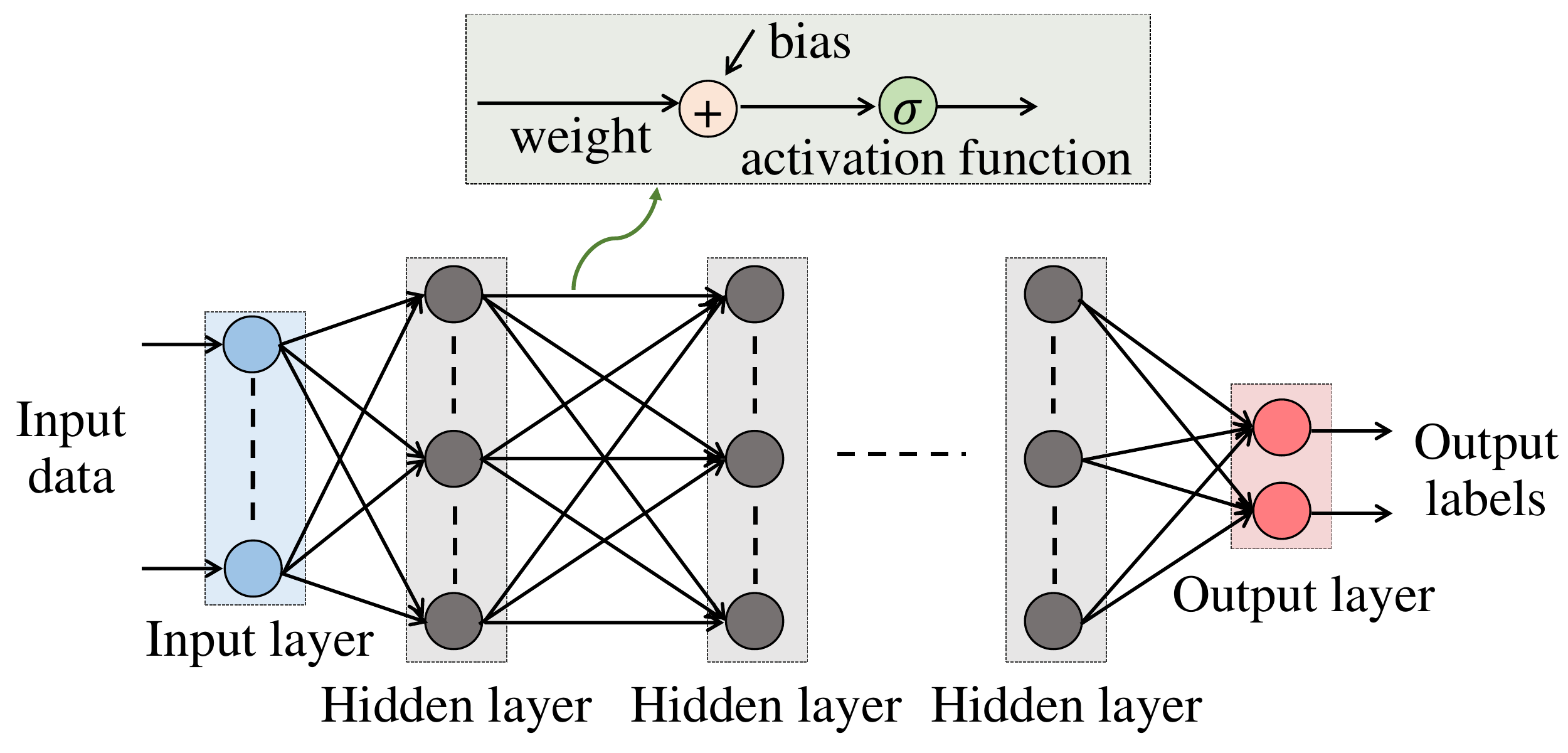} 
	\caption{The structure of a feedforward neural network.}
	\label{fig:fnn}
\end{figure}

We use TensorFlow \cite{Tensorflow} to build $\mathcal{C}_T$ with an FNN structure shown in Figure~\ref{fig:fnn}.
The following hyperparameters are selected as a local optimal solution by solving {\it OptHyper} for the deep neural network of $\mathcal{C}_T$:
\begin{itemize}
	\item An FNN is trained with backpropagation algorithm by using cross-entropy as the loss function. Cross-entropy function is given by
	\begin{eqnarray*}
{\text{CE}}(\bm{\theta}) &=& - \sum_i \left( [\bm{y}_T]_i \log [a^L(\bm{x}_T)]_i +  (1- [\bm{y}_T]_i) \cdot \right. \\
&& \left. \log (1- [a^L(\bm{x}_T)]_i) \right),
\end{eqnarray*}
where $\bm{\theta}$ is the set of the neural network parameters,
$\bm{x}_T$ is the training data vector, $\bm{y}_T$ is the corresponding label vector, and $a^L\left(\bm{x}_T\right)$ is the output of the neural network at the last (output) layer $L$.

	\item Number of hidden layers is 3.
	\item Number of neurons per hidden layer is 100.
	\item Rectified linear unit (ReLU) is used as activation function at hidden layers. ReLU performs the $f(x) = \max(0,x)$ operation on input $x$.
	\item Softmax is used as the activation function at output layer. Softmax performs $f(\bm{x})_i = {e^{x_i}}/{\sum_j e^{x_j}}$ on input $\bm{x}$.
	\item Batch size is 100.
	\item Number of training steps is 1000.
\end{itemize}

In the simulation, background traffic arrives at background transmitter $B$ at rate of $\lambda = 0.8$ packet per time slot.
When $B$ has queued data packet, it may decide to transmit at rate of $1$ packet per time slot and once it transmits, it will continue until the queue is empty.
The channel gain $g_{_{BT}}(t)$ is a random variable with a Gaussian distribution and the expected value $d_{BT}^{-2}$, where $d_{BT}$ is the distance between $B$ and $T$.
In the simulation setting, the location of $B$ is $(0,10)$, the location of $T$ is $(0,0)$, and the transmit power at $B$ is $P=1000$ (normalized with respect to the unit noise power).

$T$ collects $1000$ samples, each with the most recent $10$ spectrum sensing results and a label (`idle' or `busy'). Half of these samples are used as training data and the other half of them are used as test data.
The optimized deep learning classifier $\mathcal{C}_T$ minimizes its error $e(\mathcal{C}_T)$.
In this test phase, there are $403$ busy and $97$ idle time slots found in the test data.
Among them, $3$ busy time slots are identified as idle and no idle time slot is identified as busy.
Thus,
$e_{_{FA}}(\mathcal{C}_T) = 0\%, e_{_{MD}}(\mathcal{C}_T) = 3/403 = 0.74\%$ and
$e(\mathcal{C}_T) = 0.74\%$.
This small error shows that $T$ can reliably predict the channel status of a given time slot when there is no attack.
Note that we achieve this small error by optimizing deep learning hyperparameters.
Other parameters may result in worse performance.
For example, if the number of hidden layers is changed to $10$ and the number of neurons per hidden layer is changed to $50$, we end up with $e_{_{FA}}(\mathcal{C}_T) = 0\%, e_{_{MD}}(\mathcal{C}_T) = 5/403 = 1.24\%$, and
$e(\mathcal{C}_T) = 1.24\%$ that are worse what can be achieved with hyperparameter tuning. When we use Hyperband for hyperparameter optimization, the deep neural network is determined to have three hidden layers, with $89$, $119$, and $109$ neurons, and we end up with $e_{_{FA}}(\mathcal{C}_T) = 1.03\%, e_{_{MD}}(\mathcal{C}_T) = 0.99\%$, and
$e(\mathcal{C}_T) = 1.03\%$.

The classifier with the best set of hyperparameters is implemented on the embedded GPU platform, Nvidia Jetson Nano. The run time to get one classification result in test time (namely, to run one sample through the deep neural network) is measured as $0.31$ msec.
The run time for adversary's algorithm to be developed later is similar.

$T$ transmits in idle time slots detected by $\mathcal{C}_T$.
If the SNR (or SINR) $\gamma$ at receiver $R$ is no less than a threshold $\gamma_{min}=3$, $R$ confirms a successful transmission by sending an ACK to $T$.
We set the location of $R$ as $(10,0)$ and the transmit power at $T$ as $P=1000$ (again normalized with respect to the unit noise power).
$T$ applies its deep learning classifier on $500$ time slots and makes transmission decisions.
In this training phase, there are $404$ busy and $96$ idle time slots.
Among them, $2$ busy time slots are identified as idle and transmissions in these $2$ slots fail, while $96$ idle time slots are correctly identified as idle and transmissions in $95$ slots are successful.

We evaluate the achieved \emph{normalized throughput} $M_{\text{Th}}$, which is defined as the ratio of the number of successful transmissions to the number of idle time slots.
In simulations, we measure
$M_{\text{Th}} = 95/96=98.96\%$.
We also evaluate the \emph{success ratio} $M_{\text{Sr}}$, which is defined as the ratio of the number of successful transmissions to the number of all transmissions.
In simulations, we measure
$M_{\text{Sr}} = 95/(96+2)=96.94\%$.
Due to small errors in detecting busy/idle time slots, normalized throughput and success ratio achieved by $T$'s algorithm are high.
Finally, we evaluate the overall \emph{transmission ratio} $M_{\text{Tr}}$, which is defined as the ratio of the number of all transmissions to the number of all slots.
In simulations, we measure
$M_{\text{Tr}} = (96+2)/500 = 19.60\%$.

In this paper, we focus on deep learning based algorithms, which has better performance than other machine learning algorithms.
For example, $T$ can also use an SVM based classifier to analyze sensing data.
We found that the performance of this classifier is worse, namely $e_{FA}(\mathcal{C}_A) =  16.49\%$ and $ e_{MD}(\mathcal{C}_A) = 1.74\%$.
Both error probabilities are much larger than the performance of the deep learning classifier.
Also note that the search space of \emph{OptHyper} includes the case of one hidden layer (i.e., neural network) during the search and finds that deep learning solution with three hidden layers has higher accuracy ($99.26\%$ by deep neural network vs. $96.88\%$ by a neural network with a single hidden layer).

Next, we consider the impact of different channel models.
Three additional models, i.e., Rayleigh channel model, Rician channel model, and log-normal channel model, are studied under the same setting assumed for all other factors.
From Table~\ref{table:channel}, we can see that deep learning can build an accurate classifier for each of these channel models (with errors less than $5\%$), although the error probabilities are different under different channel models.

{\small
\begin{table}
	\caption{Results under different channel models.}
	\centering
		\begin{tabular}{c|c|c}
			Channel model & Misdetection & False alarm \\ \hline \hline
			Gaussian & $0.74\%$ & $0\%$ \\ \hline
			Rayleigh & $4.91\%$ & $4.42\%$ \\ \hline
			Rician & $2.04\%$ & $1.85\%$ \\ \hline
			log-normal & $1.03\%$ & $0.89\%$
		\end{tabular}
	\label{table:channel}
\end{table}
}

To consider the impact of locations, we change the location of background transmitter to $(0,5)$, $(0,15)$, and $(0,20)$.
Results in Table~\ref{table:location} show that error probabilities can be smaller if the background transmitter is closer to the transmitter, since the sensed signal will be stronger.

{\small
\begin{table}
	\caption{Results under different background transmitter locations.}
	\centering
		\begin{tabular}{c|c|c}
			Location & Misdetection & False alarm \\ \hline \hline
			$(0,5)$ & $0.53\%$ & $0\%$ \\ \hline
			$(0,10)$ & $0.74\%$ & $0\%$ \\ \hline
			$(0,15)$ & $0.74\%$ & $1.03\%$ \\ \hline
			$(0,20)$ & $4.08\%$ & $3.70\%$
		\end{tabular}
	\label{table:location}
\end{table}
}

We also consider the impact of mobility, i.e., the classifier is built when background transmitter is at $(0,10)$ but then it moves to $(0,5)$, $(0,15)$, and $(0,20)$, respectively.
Results in Table~\ref{table:mobility} show that error probabilities can be smaller if the background transmitter is moved closer to the transmitter, since the sensed signal will be stronger, otherwise, error probabilities will be larger as the background transmitter moves away.

{\small
\begin{table}
	\caption{Results under different background transmitter locations.}
	\centering
		\begin{tabular}{c|c|c}
			Location in test phase & Misdetection & False alarm \\ \hline \hline
			$(0,5)$ & $0.53\%$ & $0\%$ \\ \hline
			$(0,10)$ & $0.74\%$ & $0\%$ \\ \hline
			$(0,15)$ & $1.24\%$ & $1.03\%$ \\ \hline
			$(0,20)$ & $1.53\%$ & $4.63\%$
		\end{tabular}
	\label{table:mobility}
\end{table}
}

Finally, we consider the case of multiple background sources.
There are two additional background transmitters at $(-5, 10)$ and $(5,10)$ with the same transmit power as the one at $(0,10)$.
To have similar number of idle time slots, the traffic rates at all these transmitters are set as $\lambda = 0.4$ packet per time slot.
If any background transmitter is sending its data, the channel is busy.
The spectrum sensing observes the aggregated signal from all these transmitters as input to $T$'s classifier.
We find that the trained classifier has error probabilities $e_{_{FA}}(\mathcal{C}_T) = 0\%, e_{_{MD}}(\mathcal{C}_T) = 0.26\%$, i.e., it has better performance than the classifier for the case of single background source.
The reason is that multiple active sources will generate larger aggregated signal and thus it is easier to predict the channel status.

\section{Adversary's Algorithm}
\label{sec:adversary}

There is an adversary $A$ that aims to reduce transmitter $T$'s performance.
As the first step, $A$ needs to determine $T$'s time slot structure (start and end point, and duration) and its decomposition to sensing, transmission, and feedback periods. This step is discussed in detail in Section \ref{sec:scenario}.
With the knowledge of $T$'s time slot structure, $A$ launches an \emph{exploratory attack} to infer $\mathcal{C}_T$.
Then it analyzes $T$'s behavior and launches different attacks (using the same energy budget).
In this section, we consider two types of attacks.
\begin{itemize}
\item \emph{Evasion attack.} $A$ jams $T$'s sensing period such that $T$ collects wrong channel data samples and thus makes wrong decisions when it runs its classifier with these wrong samples.

\item \emph{Causative attack.} Suppose that $T$ collects additional training data and retrains $\mathcal{C}_T$.
$A$ jams $T$'s sensing period such that $T$ collects wrong training data and thus the updated classifier $\mathcal{C}_{\tilde{T}}$ fails to improve and also becomes worse.
\end{itemize}

\subsection{Exploratory Attack}

\begin{figure}
	\centering
	\includegraphics[width=1\columnwidth]{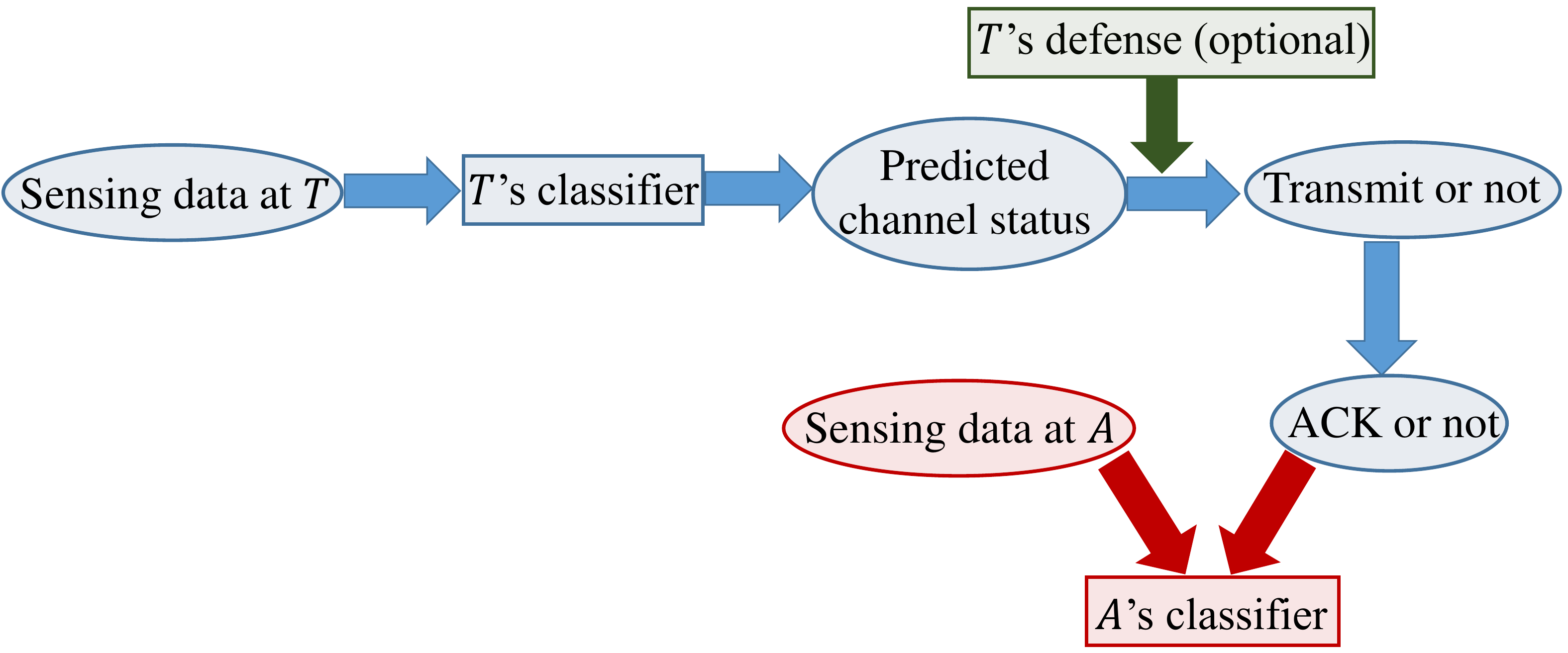} 
	\caption{The input and output (label) data while training $A$'s classifier.}
	\label{fig:asclassifiertraining}
\end{figure}

For exploratory attack, $A$ senses the spectrum, predicts whether there will be a successful transmission (if there was no attack), and performs certain attacks (if it predicts that there will be a successful transmission).
There are four cases:
\begin{enumerate}
	\item time slot $t$ is idle ($S_t=$ `idle') and $T$ is transmitting,
	\item time slot $t$ is busy ($S_t=$ `busy') and $T$ is not transmitting,
	\item time slot $t$ is  idle ($S_t=$ `idle') and $T$ is not transmitting, or
	\item time slot $t$ is busy ($S_t=$ `busy') and $T$ is transmitting.
\end{enumerate}
Since $T$ is transmitting if and only if $\mathcal{C}_T(\bm{F}_t)=$ `idle', the last two cases correspond to false alarm and misdetection of $\mathcal{C}_T$, respectively.
Our results in Section~\ref{sec:transmitter} show that these are rare cases and $\max\{ e_{MD}(\mathcal{C}_T), e_{FA}(\mathcal{C}_T) \} = 0.74\%$.
$A$ uses the most recent $n_{new}$ sensing results as the features and the current feedback (`ACK' vs. `no ACK') as the label to build one training sample.
For numerical results, $n_{new}$ is assumed to be $10$. Note that $T$ and $
A$ do not know classifier parameters of each other including $n_{new}$.
After observing a certain period of time, $A$ collects a number of samples as training data to build a deep learning classifier that outputs one of two labels, `ACK' (namely, a successful transmission) and `no ACK' (namely, a failed transmission).
Figure~\ref{fig:asclassifiertraining} shows the input data and the labels while building the adversary's classifier $\mathcal{C}_A$.
$A$'s training algorithm is summarized in Algorithm~\ref{alg:a-train}.

\begin{algorithm}[t]
    \caption{$A$'s training algorithm}
    \label{alg:a-train}
    \begin{algorithmic}[1]
        \STATE $A$ collects data over a time period to build its training data  $D_{train}$.
        \STATE $A$ builds a training sample $(\bm{F}_t, l_t)$ for each time $t \ge n_{new}$, where $\bm{F}_t = (p_{t-n_{new}-1}, p_{t-n_{new}-2}, \cdots, p_t)$, $p_t$ is the sensed power at time $t$, and $l_t$ is the label (ACK or not) at time $t$.
        \STATE $A$ trains a deep learning classifier $\mathcal{C}_A$ using its training data $D_{train}$.
    \end{algorithmic}
\end{algorithm}

The process of building $\mathcal{C}_A$ can be regarded as an \emph{exploratory attack}, since $A$ aims to build $\mathcal{C}_A$ to infer the operation of $T$.
There are the following two differences between these classifiers.
\begin{itemize}
\item Due to different locations of $T$ and $A$, and random channels, the sensing results at $T$ and $A$ differ.
Thus, features for the same sample are different at $T$ and $A$.

\item The labels (classes) at $T$ and $A$ are different, i.e., labels are `busy' or `idle' in $T$'s classifier and `ACK' or `no ACK' in $A$'s classifier.
\end{itemize}

\begin{algorithm}[t]
    \caption{$A$'s prediction algorithm}
    \label{alg:a-predict}
    \begin{algorithmic}[1]
        \STATE At time $t$, $A$ senses channel and collects received power $p_t$.
        \STATE $A$ builds a test sample $\bm{F}_t = (p_{t-n_{new}-1}, p_{t-n_{new}-2}, \cdots, p_t)$.
        \STATE $A$ uses its classifier $\mathcal{C}_A$ to decide on a label (`ACK' or `no ACK') for the test sample at time slot $t$, i.e., computes $\mathcal{C}_A(\bm{F}_t)$.
    \end{algorithmic}
\end{algorithm}

Once $\mathcal{C}_A$ is built, $A$ uses it to predict whether there is a successful transmission (if there was no attack).
This prediction algorithm is given in Algorithm~\ref{alg:a-predict}.
For this algorithm, there may be two types of errors:
\begin{itemize}
\item \emph{Misdetection}. There will be a successful transmission but $\mathcal{C}_A$ predicts that there will not be a successful transmission, i.e., $l_t =$ `ACK' and $\mathcal{C}_A(\bm{F}_t)=$ `no ACK'.

\item \emph{False alarm}. There will not be a successful transmission but $\mathcal{C}_A$ predicts that there will be a successful transmission, i.e., $l_t =$ `no ACK' and $\mathcal{C}_A(\bm{F}_t)=$ `ACK'.
\end{itemize}
$A$ aims to minimize error probability $e(\mathcal{C}_A) = \max\{e_{MD}(\mathcal{C}_A), e_{FA}(\mathcal{C}_A) \}$, where $e_{MD}(\mathcal{C}_A)$ and $e_{FA}(\mathcal{C}_A)$ are the probabilities of misdetection and false alarm for $\mathcal{C}_A$, respectively. For that purpose, it trains $\mathcal{C}_A$ and selects its hyperparameters.
The underlying optimization problem is similar to the one for $T$ (discussed in Section~\ref{sec:transmitter}) and thus its discussion is omitted here.

We use TensorFlow to build $\mathcal{C}_A$.
In the simulation, we set the location of $A$ as $(10,10)$.
$A$ collects $1000$ samples (each sample corresponds to $10$ most recent sensing results) with labels in $1000$ time slots.
Half of these samples are used as training data and the other half is used as test data.
There are $95$ successful transmissions in $500$ test data. Out of $95$ transmissions, $4$ are predicted as failed transmissions, although these transmissions are indeed successful.
Among $405$ failed transmissions, $8$ of them are predicted as successful transmissions, although these transmissions indeed fail.
Thus,
$e_{FA}(\mathcal{C}_A) = 8/405=1.98\%, e_{MD}(\mathcal{C}_A) = 4/95=4.21\%$ and $e(\mathcal{C}_A) = 4.21\%$
This small error shows that $A$ can reliably predict the successful transmissions by $T$. The inferred classifier $\mathcal{C}_A$ is further used by $A$ for two additional attacks, evasion and causative attacks, discussed next.

\subsection{Evasion Attack} \label{sec:evasion}

With $\mathcal{C}_A$, $A$ can perform an evasion attack (that targets the test time of $T$) as follows.
If $\mathcal{C}_A$ predicts that a time slot will have an ACK when there is no attack, $A$ transmits in the initial sensing period to change $T$'s sensing result for the current time slot.
This sensing result is one feature of $T$'s classifier (part of $\bm{F}_t$ in time slot $t$) and thus $T$ may make a wrong decision (namely, $T$ may misclassify the status of time slot $t$), even if $\mathcal{C}_T$ was built successfully to predict idle/busy channel states in the absence of attacks. Compared with a continuous jamming attack, this attack targets the initial sensing period that is much shorter than the data transmission period.
Hence, the power consumption of this attack is much less than continuous jamming. There are two important properties of this attack compared to jamming data transmissions. First, it is more energy-efficient and can be used to attack $T$ over a longer period of time (assuming $A$ is battery-operated). Second, it is more difficult to be detected by $T$ since $A$ does not jam transmission of $T$ (so DoS detection mechanisms cannot be readily applied).

\begin{figure}
	\centering
	\includegraphics[width=1\columnwidth]{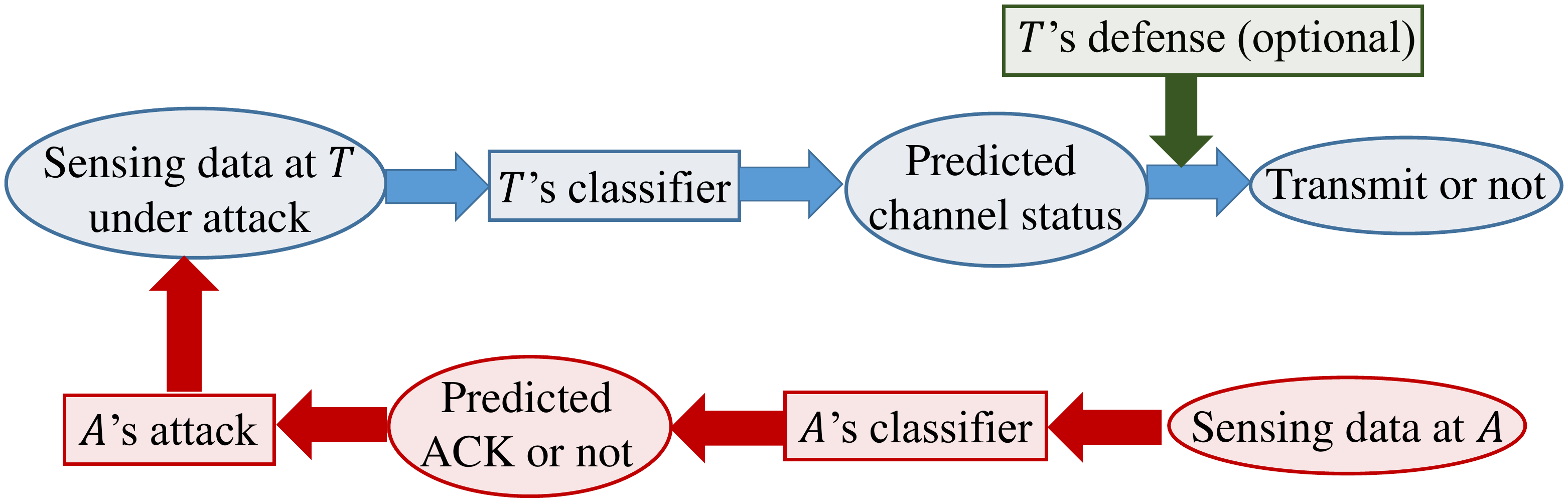}
	\caption{Using $A$'s classifier for evasion attacks.} 
	\label{fig:asclassifierruntime}
\end{figure}

Figure~\ref{fig:asclassifierruntime} illustrates $A$'s operation for evasion attack.
In the simulation, the transmit power at $A$ is set as $P=1000$.
For $\mathcal{C}_T$ built in Section~\ref{sec:transmitter} and $500$ time slots considered for transmissions under the attack,
$3$ (of $96$) idle time slots are identified as idle and the transmissions in these $3$ slots are all successful, while $1$ busy time slot is identified as idle and the transmission in this slot fails.
Thus, the achieved normalized throughput is
$M_{\text{Th}} = 3/96=3.13\%$,
and the overall success ratio is
$M_{\text{Sr}} = 3/(3+1)=75\%$,
while only very few transmission attempts are made such that the all transmission ratio is
$M_{\text{Tr}} = (3+1)/500=0.80\%$.
As a result, $A$ reduces the throughput of $T$ significantly from $M_{\text{Th}} = 98.96\%$ to $3.13\%$,
the success ratio from $M_{\text{Sr}} = 96.94\%$ to $75\%$, and the ratio of transmissions from $M_{\text{Tr}} = 19.60\%$ to $0.80\%$.

We compare the evasion attack with traditional jamming attack that targets data transmissions (as studied in \cite{Yi2018}), where $A$ jams when it predicts that $T$ may have a successful transmission (if there was no attack).
We consider the optimistic case that the prediction accuracy is the same as the deep learning classifier $\mathcal{C}_A$.
To have a fair comparison, we consider the same energy budget for these two attacks. For that purpose, we measure the energy consumption of $A$ (namely, the ratio of time slots when $A$ transmits) under the spectrum poisoning attack. Then we use this energy budget for every other attack considered in this paper.
We assume that the lengths of a sensing period and a transmission period are $10\%$ and $90\%$ of the entire time slot, respectively (we ignore the small end period for feedback).
We also assume that the energy budget allows $A$ to transmit in the entire sensing period for all time slots.
Then for jamming attack, $A$ can jam up to $1/9$ of all time slots under the same energy budget.
In this case, $A$ will select time slots with high probabilities of having an ACK if no attack.
$T$'s transmission decisions do not change, i.e., $98$ transmissions.
Among them, $40$ will be successful under jamming attack.
Given that there are $96$ idle time slots, we have $M_{\text{Th}}=40/96=41.67\%, M_{\text{Sr}}=40/98=40.82\%, M_{\text{Tr}}=98/500=19.60\%$ for the jamming attack.
We can see that under the same energy budget, jamming attack is not as effective as the evasion attack considered in this paper.

\subsection{Causative Attack}
\label{sec:causative}

$A$ can also launch a causative attack (that targets the training process of $\mathcal{C}_T$) by using $\mathcal{C}_A$ if $T$ is updating $\mathcal{C}_T$ using additional training data.
To attack the re-training process of $\mathcal{C}_T$, $A$ identifies $T$'s re-training phase, namely when it starts and ends, as follows.
We assume such re-training process is launched periodically.
Thus, $A$ can identify $T$'s re-training phase in two steps (see Algorithm~\ref{alg:retrain}).
In the first step, $A$ observes the accuracy of using $\mathcal{C}_A$ to predict ACK.
Once $T$ updates $\mathcal{C}_T$, $A$ observes a change in this accuracy.
The time instances of changes can be used to identify the time to update $\mathcal{C}_T$, which is the ending time of re-training phases.
In the second step, $A$ can launch the causative attack with adjustable length, which corresponds to different estimation on the length of a re-training phase.
If increasing this length cannot improve the impact of causative attack, the current length is no less than the length of a re-training phase.
Otherwise, the current length is no more than the length of a re-training phase.
Thus, $A$ can adjust the predicted length of re-training to determine the actual length.
The result of these two steps determines the re-training phase of $T$, as formulated in Algorithm~\ref{alg:retrain}.

\begin{algorithm}[t]
    \caption{Determine the start and end of re-training phase}
    \label{alg:retrain}
    \begin{algorithmic}[1]
        \STATE $A$ observes the accuracy of $\mathcal{C}_A$ on whether there will be an ACK and identifies time instances $t_1, t_2, \cdots, t_m$ when accuracy changes.
        \STATE $A$ finds the optimal parameter $\Delta$ such that $\sum_{i=1}^m [t_i - (t_1 + (i-1)\Delta)]^2$ is minimized, which is the time between two re-training phases.

        \STATE The initial lower and upper bounds for the re-training length are $L$ and $U$, respectively.
        \STATE $A$ starts causative attack with two initial estimated lengths $l, l+\delta$ on the re-training phase, where $l=(L+U)/2$.
        \IF{Causative attack with $l+\delta$ has the same performance as that by $l$}
            \STATE Update the upper bound as $U=l$.
        \ELSE
            \STATE Update the lower bound as $L=l$.
        \ENDIF
        \IF{$U-L \le \delta$}
            \STATE The re-training length is $L$.
        \ELSE
            \STATE Go to Step~4.
        \ENDIF
    \end{algorithmic}
\end{algorithm}

Once the re-training phases are determined, $A$ performs a causative attack by transmitting in the initial sensing period if $\mathcal{C}_A$ predicts an ACK.
To retrain $\mathcal{C}_T$,
$T$ collects additional training data but its sensing results are changed due to $A$'s transmissions.
Hence, this causative attack can change the training data and then change $T$'s classifier to $\mathcal{C}_{\tilde{T}}$.
Thus, $T$'s performance drops even if $A$ does not transmit later to change sensing results in test time.
Comparing with an evasion attack, the power consumption of this attack is even smaller than continuous jamming of sensing period.

\begin{figure}
	\centering
	\includegraphics[width=1\columnwidth]{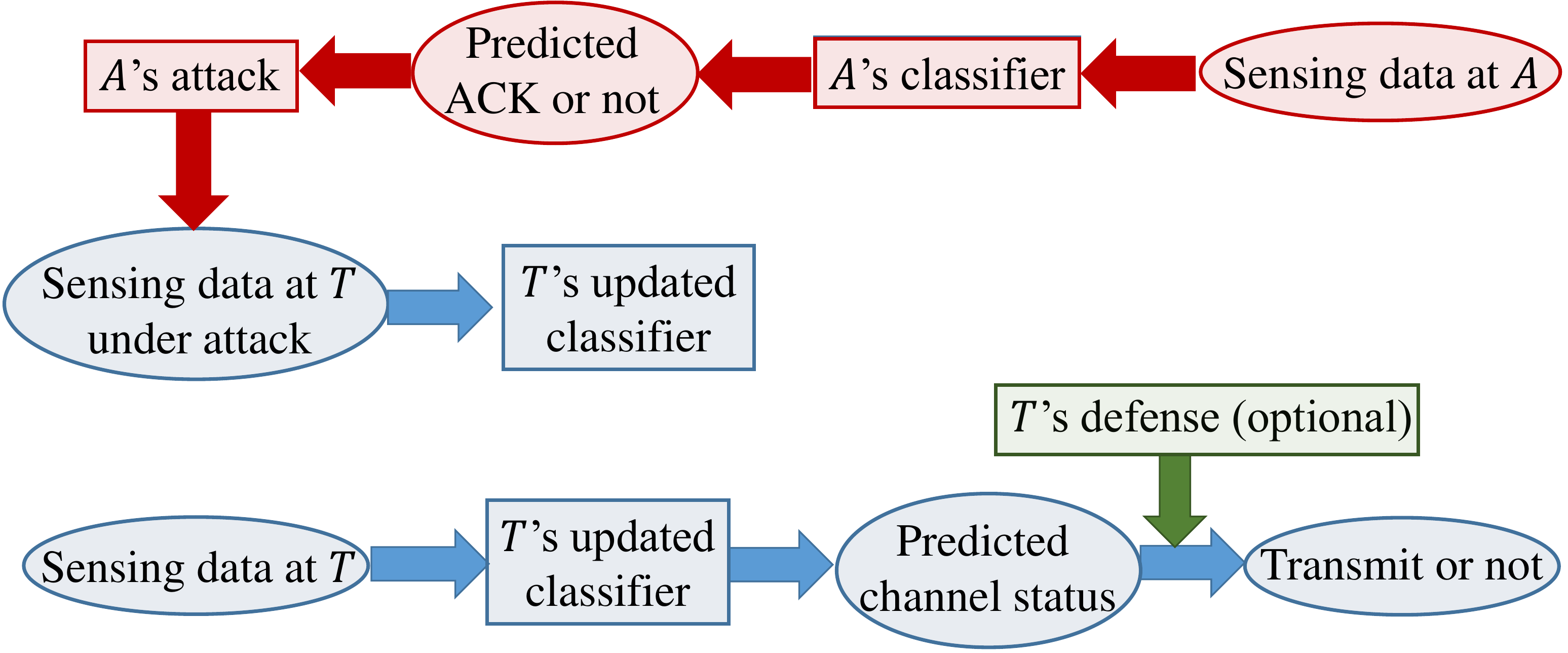} 
	\caption{Using $A$'s classifier for causative attacks.}
	\label{fig:causative}
\end{figure}

Figure~\ref{fig:causative} illustrates the adversary's operation for causative attacks.
In the simulation, the transmit power at $A$ is set as $P=1000$.
For the classifier built in Section~\ref{sec:transmitter} and $500$ time slots considered for transmissions after the attack, $89$ (of $110$) idle time slots are identified as idle and the transmissions in $88$ slots are successful, while $69$ busy time slot is identified as idle and $8$ transmissions in this slot are successful.
Thus, the achieved normalized throughput is
$M_{\text{Th}} = (88+8)/110=87.27\%$,
the overall success ratio is
$M_{\text{Sr}} = (88+8)/(89+69)=60.76\%$,
and the all transmission ratio is
$M_{\text{Tr}} = (89+69)/500=31.60\%$.
As a result, $A$ increases the ratio of transmissions from $19.60\%$ to $31.60\%$.
However, more transmissions cannot improve the performance.
$A$ reduces the throughput of $T$ from $M_{\text{Th}} = 98.96\%$ to $87.27\%$ and reduces the success ratio from $M_{\text{Sr}} = 96.94\%$ to $60.76\%$ without the need of further transmissions.

\begin{table*}
	\caption{Results under various attacks.}
	\centering
	{\small
		\begin{tabular}{c|c|c|c}
			  & Normalized & Success & All transmission \\  
			& throughput $M_{\text{Th}}$ & ratio $M_{\text{Sr}}$ & ratio $M_{\text{Tr}}$ \\ \hline \hline
			no attack & 98.96\% & 96.94\% & 19.60\% \\ \hline
			evasion attack & 3.13\% & 75.00\% & 0.80\% \\ \hline
			jamming & 41.67\% & 40.82\% & 19.60\% \\ \hline
			causative attack & 87.27\% & 60.76\% & 31.60\% \\ \hline
			causative + evasion attack & 2.72\% & 75.00\% & 0.80\% \\ \hline
			causative + jamming attack & 37.27\% & 25.95\% & 31.60\%
		\end{tabular}
	}
	\label{table:attack}
\end{table*}

\subsection{Causative Attack followed by Evasion or Jamming Attack}
\label{sec:combined}

The causative attack can be followed by an evasion attack.
That is, $A$ first launches the causative attack such that $T$'s classifier is updated as $\mathcal{C}_{\tilde{T}}$ with wrong samples of additional training data.
Then $A$ also launches the evasion attack such that the input features to $\mathcal{C}_{\tilde{T}}$ are also wrong.
As a result, $A$ reduces the throughput of $T$ from $M_{\text{Th}} = 98.96\%$ to $2.72\%$, the success ratio from $M_{\text{Sr}} = 96.94\%$ to $75.00\%$, and the ratio of transmissions from $M_{\text{Tr}} = 19.60\%$ to $0.80\%$.

The causative attack can also be followed by a jamming attack (that targets data transmissions) with an energy budget.
As discussed in Section~\ref{sec:evasion}, we assume that $A$ can jam up to $1/9$ of all time slots.
Under this setting, $A$ increases the ratio of transmissions from $19.60\%$ to $31.60\%$.
Again, more transmissions cannot improve the performance.
$A$ reduces the throughput of $T$ from $M_{\text{Th}} = 98.96\%$ to $5.45\%$ and reduces the success ratio from $M_{\text{Sr}} = 96.94\%$ to $3.80\%$.

Table~\ref{table:attack} summarizes the performance of $T$ without an attack and with various attacks considered in this paper, and demonstrates the success of these attacks. Overall, the proposed attacks cause major loss in $T$' performance and the impact is much more substantial than typical jamming attacks that target data transmissions under the same energy budget.

\section{Defense Strategy}
\label{sec:defense}

The first step of the proposed attacks is an exploratory attack to understand how $\mathcal{C}_T$ works and build $\mathcal{C}_T$.
One approach to protect a deep learning algorithm against attacks is adding some randomness to the deep neural network of the target and making it more challenging for the adversary to learn its structure \cite{Kurakin18:defense}.
However, this approach is not effective, since $A$ does not have access to the last layer of the neural network of $T$. However, $A$ can access the outcome (ACK or not) of $T$'s actions (transmissions).
Therefore, an alternative approach is to add randomness directly to $T$'s transmissions, which will in turn change the input to $A$ (namely, the labels collected by $A$ to build its classifier in the exploratory attack).
Note that a small level of randomness may not change ACKs much and thus $A$ can still perform an exploratory attack.
On the other hand, a large level of randomness will randomly change $T$'s actions, which makes $T$'s performance worse, even without attack.
Note that we consider a single channel system and thus spectrum handoff to other channels is not possible as a strategy to confuse the adversarial attack.

In this paper we consider a defense strategy that selectively changes $T$'s actions, i.e., makes $T$ transmit in a time slot when it is identified as busy\footnote{If there is an attack, this defense will improve the performance as we show later in this section. However, the impact of this defense action on throughput is not obvious if there are multiple transmitters but there is no attack. The reason is that $\mathcal{C}_T$ is not perfect and thus an idle channel may be identified as busy. This issue can be resolved by an alternative defense at $R$ that sends ACK although no packet is received. As discussed later in this section, this approach achieves the same defense performance on average without issue of additional interference to other nodes.} or not transmit in a time slot when it is identified as idle.
Such changes should ensure that the observation of $A$ becomes incorrect and thus it cannot build a good classifier $\mathcal{C}_A$.
As a consequence, $A$ cannot perform subsequent attacks effectively, as well.
Moreover, such changes should not reduce $T$'s performance significantly.
Therefore, this defense mechanism involves a fundamental trade-off  between the accuracy of $\mathcal{C}_A$ and the performance of $T$ after taking some defense actions.
The problem is how to select a number of time slots such that taking defense actions on these slots can achieve the maximum (negative) impact on the accuracy of $A$'s classifier.
This can be formulated as an optimization problem as follows.
\begin{eqnarray*}
\mbox{{\it OptDefense}:} && \\
\mbox{maximize} && e(\mathcal{C}_A) \\
\mbox{subject to} && e(\mathcal{C}_A) \le e_{MD}(\mathcal{C}_A), \; e(\mathcal{C}_A) \le e_{FA}(\mathcal{C}_A), \\
&& \hspace{-5mm} e_{MD}(\mathcal{C}_A) = \frac{n_{MD}(\mathcal{C}_A)}{n_{busy}}, \; e_{FA}(\mathcal{C}_A) = \frac{n_{FA}(\mathcal{C}_A)}{n_{idle}}, \\
&& \mathcal{C}_A = \mathcal{L}_A(H, \hat{D}_{train}, \hat{D}_{test}), \; H \in \mathcal{H}, \\
&& \frac{|D_{train} - \hat{D}_{train}|}{|D_{train}|} \le P_d,
\end{eqnarray*}
where $\hat{D}_{train}$ and $\hat{D}_{test}$ are training and test data sets for $A$ under $T$'s defense actions, the subtraction of sets $D_{train} - \hat{D}_{train}$ is the set of elements in $D_{train}$ but not in $\hat{D}_{train}$, $|D|$ denotes the size of set $D$, $\frac{|D_{train} - \hat{D}_{train}|}{|D_{train}|}$ is the ratio of defense actions over $D_{train}$, and $P_d$ is the maximum allowed ratio on defense actions. $A$'s function for deep learning process, $\mathcal{L}_A$, depends not only on $H$ and $\hat{D}_{train}$ but also on $\hat{D}_{test})$, since $T$ does not know when $A$ collects training data and when $A$ collects test data. Thus, $T$ takes some ratio of defense actions on $D_{train}$ and $D_{test}$ (assumed to be equal for numerical results).

\begin{table*}
\caption{Results for defense strategy under evasion attack.}
\centering
{\small
\begin{tabular}{c|c|c|c|c}
\# of defense operations & \multicolumn{2}{|c|}{Adversary error probabilities} & \multicolumn{2}{|c}{Transmitter performance} \\ \cline{2-5}
divided by \# of all samples & Misdetection & False alarm & Normalized throughput & Success ratio \\ \hline \hline
0\% (no defense) & 1.98\% & 4.21\% & 3.13\% & 75.00\% \\ \hline
10\% & 6.99\% & 10.59\% & 15.63\% & 15.31\% \\ \hline
20\% & 8.92\% & 35.29\% & 41.67\% & 28.78\% \\ \hline
40\% & 10.12\% & 42.67\% & 51.04\% & 18.22\% \\ \hline
60\% & 17.06\% & 69.44\% & 76.04\% & 18.07\% \\ \hline
80\% & 10.88\% & 93.22\% & 56.25\% & 13.30\% \\ \hline
\end{tabular}
}
\label{table:defense}
\end{table*}

{\it OptDefense} is solved by analyzing the output of $\mathcal{C}_T$ as follows. $\mathcal{C}_T$ provides not only a label for each sample, but also a score that can be used to measure the confidence of this classification.
That is, there is a score $p(s) \in [0,1]$ for each sample $s$.
Classifier $\mathcal{C}_T$ uses a decision boundary for classification, i.e.,
if $p(s)$ is less than some decision boundary $\tau$, sample $s$ is classified as idle, otherwise, sample $s$ is classified as busy.
Note that $\tau$ is a hyperparameter in deep learning and it is selected (along with other hyperparameters) to minimize $e(\mathcal{C}_T)$.
If the difference between $p(s)$  and $\tau$, namely $|p(s) - \tau|$, is large, the confidence of the classification is high; otherwise, the confidence of the classification is low.
Therefore, to maximize the impact of defense actions, $T$ should select time slots (samples) with scores far away from the 
decision boundary.
This decision algorithm with defense is summarized in Algorithm~\ref{alg:t-defense}.
Thresholds $\tau_0$ and $\tau_1$ in Step 4 can be determined by using $D_{test}$.\footnote{Alternatively, these thresholds can also be determined by $D_{train}$. We use $D_{test}$ to be consistent with other results in the paper, i.e., we always use a classifier on test data to obtain performance results.}
For example, we can select $P_d$ of time slots by selecting $\tau_0$ such that $|\{ s : p(s) < \tau_0\}| = P_d \cdot |\{ s : p(s) < \tau\}|$ and selecting $\tau_1$ such that $|\{ s : p(s) > \tau_1\}| = P_d \cdot |\{ s : p(s) > \tau\}|$.
The probability in Step 5 is to randomize $T$'s defense actions among selected time slots, which makes $A$'s learning more challenging.
\begin{algorithm}[t]
    \caption{$T$'s defense algorithm}
    \label{alg:t-defense}
    \begin{algorithmic}[1]
        \STATE At time $t$, $T$ senses channel and obtains power $p_t$.
        \STATE $T$ builds a sample $s$ with features $\bm{F}_t = (p_{t-n_{new}-1}, p_{t-n_{new}-2}, \cdots, p_t)$.
        \STATE $T$ uses its classifier $\mathcal{C}_T$ to decide on a label (busy or idle) and a score $p(s)$.
        \IF{$p(s) < \tau_0$ or $p(s) > \tau_1$}
            \STATE $T$ changes the label with certain probability.
        \ENDIF
        \STATE $T$ transmits if time slot $t$ is still classified as idle, i.e., $\mathcal{C}_T(\bm{F}(t)) =$ `idle'.
    \end{algorithmic}
\end{algorithm}

Table~\ref{table:defense} shows the results for different $P_d$ under an evasion attack.
With more frequent defense actions (larger $P_d$), the achieved normalized throughput $M_{\text{Th}}$ increases from $3.13\%$ to $76.04\%$.
However, further increases in $P_d$ (beyond $60 \%$) reduce $M_{\text{Th}}$, as the $T$'s own channel access becomes excessively unreliable.
Results for defense strategy against causative attack are similar and thus are omitted.
We can design a search process for $P_d$ to maximize throughput, according to the adversary's actions.

The searching process of $P_d$ also works when there is no adversary.
For this extreme case, $T$ will find that any defense actions will decrease the throughput and thus the search process will end with the defense level $0\%$ without knowing whether an adversary is present, or not.

The above defense is performed by $T$.
Instead, $R$ can also perform defense actions to fool the adversary, i.e., $R$ can refrain from sending an ACK when a packet is received or send an ACK when no packet is received.
Then $A$ will observe incorrect labels to build $\mathcal{C}_A$.
$R$'s defense strategy can be realized to have the same outcome (ACK or no ACK) as $T$'s strategy by using 1 bit overhead.
There are the following three cases for $T$'s defense strategy.
\begin{itemize}
\item Case I: $T$ takes a defense action of not transmitting when channel is detected as idle. Then $R$ will not send ACK since there is no transmission.

\item Case II: $T$ takes a defense action of transmitting when channel is detected as busy. Then it is likely that $R$ will not send ACK since transmission may fail. But if such a transmission is successful, $R$ will send ACK.

\item Case III: $T$ does not take a defense action. Then $R$ will send ACK if there is a successful transmission.
\end{itemize}
To ensure the same outcome, $R$'s defense strategy is implemented for the above three cases as follows.
\begin{itemize}
\item Case I: $T$ transmits data with a 1 bit flag of ``defense action". $R$ will not send ACK even if the transmission is successful.

\item Case II: $T$ transmits 1 bit flag of ``defense action". $R$ will send ACK if the flag is successfully received.

\item Case III: If $T$ transmits data, $T$ also transmits a 1 bit flag of ``no defense action". $R$ will send ACK if the transmission is successful.
\end{itemize}
From adversary's point of view, the two defense strategies will provide the same outcome and thus the adversary will build the same classifier under the exploratory attack.
The performance under subsequent attacks (discussed in Sections~\ref{sec:evasion}, \ref{sec:causative} and \ref{sec:combined}) will be similar.
The only difference is that under $R$'s defense strategy, there are more transmissions (for Case I) and thus throughput can be further improved if these transmissions are successful.

\section{Conclusion}
\label{sec:conclusion}
We applied adversarial machine learning (based on deep neural networks) to design over-the-air spectrum sensing poisoning attacks that target the spectrum sensing period  and manipulate the input data of the transmitter in test and training phases (in form of evasion and causative attacks). An adversary launches these attacks either to fool the transmitter into making wrong transmit decisions (namely, an evasion attack) or manipulate its retraining process (namely, a causative attack). Since the adversary only needs to transmit for a short period of time to manipulate the transmit decisions, these attacks more energy-efficient and harder to detect compared to directly jamming data transmissions. We showed that these attacks substantially decrease the throughput of the transmitter and are more effective than conventional jamming attacks. We also combined evasion, causative, and jamming attacks, and measured their total impact.
To mitigate these attacks, we developed an effective defense strategy for the transmitter that intentionally takes wrong actions in selected time slots to mislead the adversary. These time slots are selected from those with the classification score that is far away from the decision boundary.
We showed that the proposed defense mechanism significantly increases the errors in adversary's decisions and prevents major losses in the performance of the transmitter.

\section*{Acknowledgements}
A preliminary version of the material in this paper was partially presented at IEEE Military Communications Conference (MILCOM), 2018. This effort is supported by the U.S. Army Research Office under contract W911NF-17-C-0090. The content of the information does not necessarily reflect the position or the policy of the U.S. Government, and no official endorsement should be inferred.


\begin{thebibliography}{99}

\bibitem{Clancy2007}
C. Clancy, H. J. Stuntebeck, and T. O'Shea, ``Applications of machine learning to cognitive radio networks," \emph{IEEE Wireless Communications}, 2007.

\bibitem{Chen2017}
M. Chen, U. Challita, W. Saad, C. Yin, and M. Debbah, ``Machine Learning for Wireless Networks with Artificial Intelligence: A Tutorial on Neural Networks,'' \emph{arXiv preprint arXiv:1710.02913}, 2017.

\bibitem{Simeone}
O. Simeone, ``A very short introduction to machine learning with applications to communication systems," \emph{IEEE Transactions Cognitive Communications and Networking}, 2018

\bibitem{Lee2017}
W. Lee, M. Kim, D. Cho, and R. Schober, ``Deep Sensing: Cooperative Spectrum Sensing Based on Convolutional Neural Networks,'' \emph{arXiv preprint arXiv:1705.08164}, 2017.

\bibitem{DeepOFDM}H. Ye, G. Y. Li, and B.-H. Juang, ``Power of Deep Learning for Channel Estimation and Signal Detection in OFDM Systems,'' \emph{IEEE Wireless Communications Letters}, 2018

\bibitem{Yi2018}
Y. Shi, Y. E Sagduyu, T. Erpek, K. Davaslioglu, Z. Lu, and J. Li,
``Adversarial deep learning for cognitive radio security: Jamming attack and defense strategies,''
\emph{IEEE International Conference on Communications (ICC) Workshop on Promises and Challenges of Machine Learning in Communication Networks}, 2018.

\bibitem{Tugba2018}
T. Erpek, Y. E. Sagduyu, and Y. Shi,
``Deep learning for launching and mitigating wireless jamming attacks," \emph{IEEE Transactions on Cognitive Communications and Networking}, 2019.

\bibitem{OShea2016}
T. O'Shea, J. Corgan, and C. Clancy, ``Convolutional radio modulation recognition networks,"  \emph{International Conference on Engineering Applications of Neural Networks}, 2016.

\bibitem{Kemal2018}
K. Davaslioglu and Y. E. Sagduyu,
``Generative adversarial learning for spectrum sensing,"
\emph{IEEE International Conference on Communications (ICC)},  2018.


\bibitem{Trappe}
W. Xu, W. Trappe, Y. Zhang, and T. Wood, ``The Feasibility of Launching and Detecting Jamming Attacks in Wireless Networks," ACM International Symposium on Mobile Ad Hoc Networking and Computing (Mobihoc05), 2005.

\bibitem{Zou15:eavesdropping}
Y. Zou, J. Zhu, L. Yang, Y.-C. Liang, and Y.-D. Yao,
``Securing physical-layer communications for cognitive radio networks,"
\emph{IEEE Communications Magazine}, 2015.

\bibitem{Clancy08:CogSec}
T. C. Clancy, and N. Goergen,
``Security in cognitive radio networks: Threats and mitigation,"
\emph{IEEE Conference on Cognitive Radio Oriented Wireless Networks and Communications (CrownCom)},
2008.

\bibitem{PUE}
Z. Yuan and D. Niyato and H. Li and J. B. Song and Z. Han, ``Defeating
primary user emulation attacks using belief propagation in cognitive
radio networks," \emph{IEEE Journal Selected Areas in Communications}, 2012.

\bibitem{Bian2008}
R. Chen, J. Park, and K. Bian, ``Robust distributed spectrum sensing in cognitive radio networks," \emph{IEEE Conference on Computer Communications (INFOCOM)}, 2008.



\bibitem{Fredrikson}
M. Fredrikson, S. Jha, and T. Ristenpart,
``Model inversion attacks that exploit confidence information and basic countermeasures,"
\emph{ACM SIGSAC Conference on Computer and Communications Security}, 2015.


\bibitem{Kurakin}
A. Kurakin, I. Goodfellow, and S. Bengio,
``Adversarial examples in the physical world,"
\emph{arXiv preprint arXiv:1607.02533}, 2016.

\bibitem{Biggio}
B. Biggio, I. Corona, D. Maiorca, B. Nelson, N. Srndic, P. Laskov, G. Giacinto, and F. Roli,
``Evasion attacks against machine learning at test time,"
\emph{European Conference on Machine Learning and Principles and Practice of Knowledge Discovery in Databases}, 2013.

\bibitem{Biggio2}
B. Biggio, B. Nelson, and P. Laskov, ``Poisoning attacks against support vector machines," \emph{International Conference on International Conference on Machine Learning}, 2012.



\bibitem{DOS}
Y. E. Sagduyu and A. Ephremides,
``A game-theoretic analysis of denial of service attacks in wireless random access,"
\emph{Journal of Wireless Networks}, 2009.

\bibitem{Kurakin18:defense}
A.~Kurakin, et al.,
``Adversarial attacks and defences competition,"
\emph{arXiv preprint arXiv:1804.00097}, 2018.

\bibitem{Ateniese}
G. Ateniese, L. Mancini, A. Spognardi, A. Villani, D. Vitali, and G. Felici,
``Hacking smart machines with smarter ones: How to extract meaningful data from machine learning classifiers," \emph{International Journal of Security and Networks}, 2015.

\bibitem{Tramer}
F. Tramer, F. Zhang, A. Juels, M. Reiter, and T. Ristenpart,
``Stealing machine learning models via prediction APIs,"
\emph{USENIX Security}, 2016.



\bibitem{HST2018}
Y. Shi, Y. E. Sagduyu, K. Davaslioglu, and J. Li, ``Active Deep Learning Attacks under Strict Rate Limitations for Online API Calls," \emph{IEEE Symposium on Technologies for Homeland Security}, 2018.




\bibitem{LarssonAML}
M. Sadeghi and E. G. Larsson, ``Adversarial attacks on deep-learning based radio signal classification," \emph{IEEE Wireless Communications Letters}, 2018.

\bibitem{Headley19}
B. Flowers, R. M. Buehrer, and W. C. Headley, ``Evaluating adversarial evasion attacks in the
context of wireless communications," \emph{arXiv preprint, arXiv:1903.01563}, 2019.

\bibitem{Deniz19}
M. Z. Hameed, A. Gyorgy, and D. Gunduz, ``Communication without interception: defense against deep-learning-based modulation detection," \emph{arXiv preprint, arXiv:1902.10674}, 2019.

\bibitem{Silvija19}
S. Kokalj-Filipovic and R. Miller, ``Adversarial examples in RF deep learning: detection of the attack and its physical robustness," \emph{arXiv preprint, arXiv:1902.06044}, 2019.

\bibitem{Tansu2018}
S. Weerasinghe, T.Alpcan, S. M. Erfani, C.Leckie, P. Pourbeik, and J. Riddle, ``Deep learning based game-theoretical approach to evade jamming attacks," \emph{International Conference on Decision and Game Theory for Security (GameSec)}, 2018.

\bibitem{Shi18:poisoning}
Y.~Shi, T.~Erpek, Y. E.~Sagduyu, and J. H.~Li,
``Spectrum data poisoning with adversarial deep learning," in \emph{Proc.~IEEE Military Communications Conference (MILCOM)}, 2018.

\bibitem{Zou15:SecurePhy}
Y.~Zou, J.~Zhu, L.~Yang, Y.~Liang, and Y.~Yao,
``Securing physical-layer communications for cognitive radio networks,''
\emph{IEEE Communications Magazine}, 2015.

\bibitem{Penna}
F. Penna, Y. Sun, L. Dolecek, and D. Cabric, ``Detecting and counteracting statistical attacks in cooperative spectrum sensing," \emph{IEEE Transactions on Signal Processing}, 2012.

\bibitem{Yut}
F. R. Yut, H. Tang, M. Huang, Z. Lit, and P. C. Mason,
``Defense against spectrum sensing data falsification attacks in mobile ad hoc networks with cognitive radios,''
\emph{IEEE Military Communications Conference (MILCOM)}, 2009.

\bibitem{Sagduyu11:jamming}
Y. E. Sagduyu, R. Berry, and A. Ephremides,
``Jamming games in wireless networks with incomplete information,"
\emph{IEEE Communications Magazine}, 2011.


\bibitem{Sagduyu09:noncoop}
Y. E. Sagduyu, R. Berry, and A. Ephremides,
``MAC games for distributed wireless network security with incomplete information of selfish and malicious user types," \emph{IEEE International Conference on Game Theory for Networks (GameNets)},
2009.

\bibitem{Lu2017}
Z. Lu, Y. E. Sagduyu, and J. Li,
``Securing the backpressure algorithm for wireless networks,"
\emph{IEEE Transactions on Mobile Computing}, 2017.

\bibitem{LuCliff2017}
Z. Lu and C. Wang, ``Enabling network anti-inference via proactive strategies: a fundamental perspective," \emph{IEEE/ACM Transactions on Networking}, 2017.



\bibitem{Saad2018}
A. Ferdowsi and W. Saad, ``Deep learning for signal authentication and security in massive internet of things systems," \emph{arXiv preprint arXiv:1803.00916}, 2018.

\bibitem{TwoDim} G. Han, L. Xiao, and H. V. Poor, ``Two-dimensional anti-jamming communication based on deep reinforcement learning,'' \emph{IEEE International Conference on Acoustics, Speech and Signal Processing (ICASSP)}, 2017.

\bibitem{Wu2017}
Z. Wu. Y. Zhao, Z. Yin, and H. Luo, ``Jamming signals classification using convolutional neural network," \emph{IEEE International Symposium on Signal Processing and Information Technology (ISSPIT)}, 2017.

\bibitem{gunes}
O. A. Topal, S. Gecgel, E. M. Eksioglu, and G. Karabulut Kurt, ``Identification of smart jammers: Learning based approaches using wavelet representation," \emph{arXiv preprint, arXiv:1901.09424}, 2019.



\bibitem{Poor2018}
L. Xiao, D. Jiang, D. Xu, H. Zhu, Y. Zhang, and V. Poor, ``Two-dimensional anti-jamming mobile communication based on reinforcement learning," \emph{IEEE Transactions on Vehicular Technology}, 2018.
\bibitem{Xiao18}
L.~Xiao, C.~Xie, M.~Min, and W.~Zhuang, ``User-centric view of unmanned aerial vehicle transmission against smart attacks," \emph{IEEE Transactions on Vehicular Technology}, 2018.

\bibitem{Liang18}
Y.~Liang, Z.~Cai, J.~Yu, Q.~Han, and Y.~Li, ``Deep learning based inference of private information using embedded sensors in smart devices," \emph{IEEE Network}, 2018.

\bibitem{Bergstra12:parameter}
J.~Bergstra and Y.~Bengio,``Random search for hyper-parameter optimization,"
\emph{Journal of Machine Learning Research}, 2012.
\bibitem{Tensorflow}
M. Abadi, \emph{et al.}, ``TensorFlow: Large-scale machine learning on heterogeneous systems," 2015. www.tensorflow.org


\bibitem{Xiao18:crowdsensing}
L.~Xiao, D.~Jiang, D.~Xu, W.~Su, N.~An, and D.~Wang, ``Secure mobile crowdsensing based on deep learning," \emph{China Communications}, vol.~15, no.~10, pp.~1--11, 2018.


\bibitem{Xiao18:antijam}
L.~Xiao, X.~Wan, W.~Su, and Y.~Tang, ``Anti-jamming underwater transmission with mobility and learning," \emph{IEEE Communications Letters}, vol.~22, no.~3, pp.~542--545, 2018.

\bibitem{Li16}
L.~Li, K.~Jamieson, G.~DeSalvo, A.~Rostamizadeh, A.~Talwalkar, ``Hyperband: A novel bandit-based approach to hyperparameter optimization," \emph{arXiv preprint, arXiv:1603.06560}, 2016.
\end{thebibliography}
\end{document}